\documentclass[modern]{aastex631}

\usepackage[T1]{fontenc}
\usepackage{soul}
\usepackage{mathtools}

\shorttitle{In icy worlds, size matters!}
\shortauthors{Kang et al.}
\graphicspath{{./}{figures/}}

\begin{document}

\title{In icy ocean worlds, size matters!}

\correspondingauthor{Wanying Kang}
\email{wanying@mit.edu}

\author[0000-0002-4615-3702]{Wanying Kang}
\affiliation{Earth, Atmospheric and Planetary Science Department, 
  Massachusetts Institute of Technology,
  Cambridge, MA 02139, USA}

\author{Malte Jansen}
\affiliation{Department of the Geophysical Sciences, University of Chicago, Chicago, Illinois}



\begin{abstract}
The ice shell and subsurface ocean on icy worlds are strongly coupled together -- heat and salinity flux from the ice shell induced by the ice thickness gradient drives circulation in the ocean, and in turn, the heat transport by ocean circulation shapes the ice shell. Since measurements in the near future are likely to remain constrained to above the ice shell, understanding this ocean-ice interaction is crucial. Using an ocean box model and a series of experiments simulating the 2D ocean circulation, we find that large icy moons with strong gravity tend to have stronger ocean heat transport under the same ice-shell topography. As a result, the equilibrium ice shell geometry is expected to be flatter on moons with larger size, and vice versa. This finding is broadly consistent with the observed ice shell geometry for Enceladus and Europa.
\end{abstract}



\section{Introduction}
Many of the icy satellites in the outer solar system are likely to contain a subsurface ocean underneath their ice shell due to tidal dissipation \citep{Scharf-2006:potential}, which may lead to a suitable environment for life to thrive. Enceladus (a satellite of Saturn) and Europa (a satellite of Jupiter), in particular, have been confirmed to have a global subsurface ocean, thanks to the Galileo and Cassini missions \citep{Postberg-Kempf-Schmidt-et-al-2009:sodium, Thomas-Tajeddine-Tiscareno-et-al-2016:enceladus, Carr-Belton-Chapman-et-al-1998:evidence, Kivelson-Khurana-Russell-et-al-2000:galileo, Hand-Chyba-2007:empirical}. Ongoing geological activities send samples of the ocean to outer space as geyser-like jets, intermittently on Europa \citep{Roth-Saur-Retherford-et-al-2014:transient, Jia-Kivelson-Khurana-et-al-2018:evidence, Arnold-Liuzzo-Simon-2019:magnetic, Huybrighs-Roussos-Bloecker-et-al-2020:active} and continuously on Enceladus \citep{Porco-Helfenstein-Thomas-et-al-2006:cassini, Hansen-Esposito-Stewart-et-al-2006:enceladus, Howett-Spencer-Pearl-et-al-2011:high, Spencer-Howett-Verbiscer-et-al-2013:enceladus}, providing a unique opportunity to peek through their kilometers thick ice shells.

As some of the most enigmatic targets to search for extraterrestrial life \citep{Des-Nuth-Allamandola-et-al-2008:nasa, Hendrix-Hurford-Barge-et-al-2019:nasa}, icy moons are to be thoroughly explored in the future (e.g., \textit{Europa Clipper} and \textit{JUICE}). However, measurements and detection are likely to be carried out above the ice shell, for the most part, due to the high cost of a drill mission. These measurements (libration, shape, gravity etc.) will allow us to get a good estimate of the icy moons' ice shell geometry. Thus far, Enceladus' ice shell has been revealed to be around 20~km thick on global average and to significantly thin toward the poles \citep{Iess-Stevenson-Parisi-et-al-2014:gravity, Beuthe-Rivoldini-Trinh-2016:enceladuss, Tajeddine-Soderlund-Thomas-et-al-2017:true, Cadek-Soucek-Behounkova-et-al-2019:long, Hemingway-Mittal-2019:enceladuss}. Europa's ice shell geometry is not as well constrained, but evidence has been found in favor a relatively thin \citep[$<$15~km][]{Hand-Chyba-2007:empirical} and flat \citep{Nimmo-Thomas-Pappalardo-et-al-2007:global} ice shell.

The ice geometry may be able to inform us of the properties of the subsurface ocean if we understand how the ice shell interacts with the subsurface ocean. As illustrated by \citet{Kang-Mittal-Bire-et-al-2021:how}, the interaction happens in a mutual way: the variation of ice thickness on Enceladus can drive ocean circulation by inducing salinity flux through freezing/melting and by changing the local freezing point; in turn, ocean circulation can converge heat to regions covered by relatively thick ice, flattening the ice shell (sketched in Fig.~\ref{fig:EOS-Hice-Heatflux}d). If the ice shell is in equilibrium, freezing and melting should be in balance with the ice flow from the thick-ice regions to the thin-ice regions \citep{Ashkenazy-Sayag-Tziperman-2018:dynamics, Kang-Flierl-2020:spontaneous}. Given the observed ice geometry \citep{Hemingway-Mittal-2019:enceladuss}, one may be able to make inferences about the ocean salinity, the partition of heat production between the ice shell and the silicate core and the ocean dissipation, because they all have impacts on the ocean circulation and heat transport. Similarly, knowing the ocean parameters, one would also be able to predict how the equilibrium ice shell geometry should look like.


Since size and gravity vary from case to case, first and foremost, we need to investigate how the size of an icy satellite affects the direction and strength of the circulation and the associated heat transport.
In the rest of this paper, we will use a conceptual model \citep[similar to that used by Stommel][]{Marotzke-2000:abrupt} to demonstrate that 1) heat transport increases with the moon's radius to the cubic power, and 2) the equilibrium equator-to-pole ice thickness difference decreases as the moon gets bigger. Numerical models will be carried out to support these scaling laws.
As it turns out, this exercise leads to methods that either allow us to constrain the ocean properties using ice geometry or to predict the ice thickness variation, depending on the size of the moon.

\section{Why would size matter?}
\label{sec:size-effect}

\begin{figure*}
    \centering \includegraphics[page=1,width=0.9\textwidth]{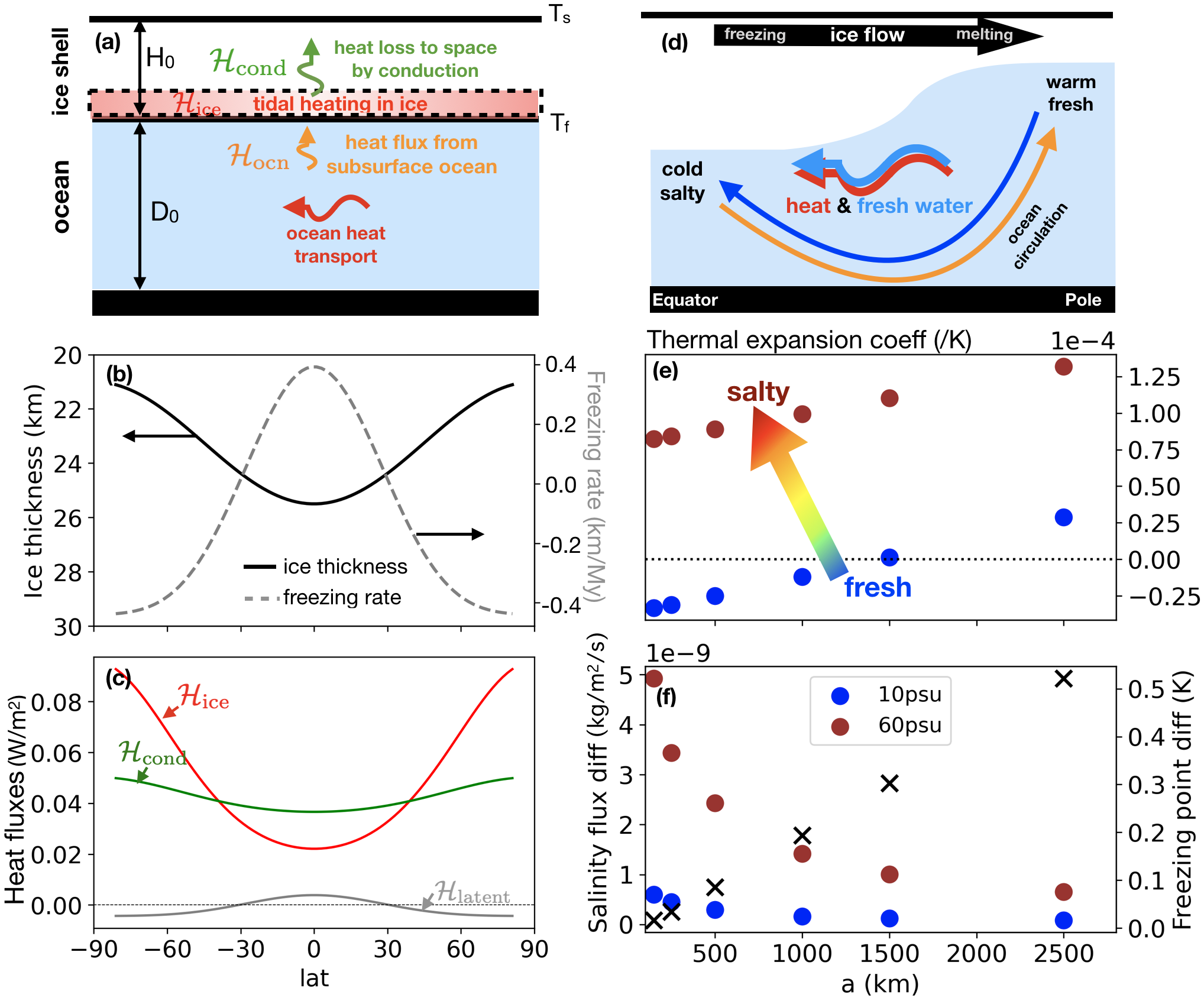}
    \caption{\small{Panel (a) defines the primary sources of heat and heat fluxes assumed here, which include: heating due to tidal dissipation in the ice $\mathcal{H}_{\mathrm{ice}}$, the heat flux from the ocean to the ice $\mathcal{H}_{\mathrm{ocn}}$ and the conductive heat loss to space $\mathcal{H}_{\mathrm{cond}}$. Ocean heat transport is shown by the horizontal arrow. Panel (b) shows the assumed ice shell thickness profile with a black solid curve, which is thinner over the poles because ice dissipation amplifies going poleward \citep{Beuthe-2019:enceladuss}. The gray dashed curve shows the freezing (positive) and melting rate (negative) required to maintain a steady state based on an upside-down shallow ice flow model (see appendix for details). In this calculation, 250~km radius is assumed. Panel (c) shows the profiles of $\mathcal{H}_{\mathrm{ice}}$, $\mathcal{H}_{\mathrm{cond}}$ and $\mathcal{H}_{\mathrm{latent}}$ given the information in panel (b). Panel (d) sketches the key physical processes in an ocean covered by an ice shell with varying thickness (see main text for description). Panel (e) shows how thermal expansion coefficient under the ice shell varies with the satellite's size (gravity), assuming 10~psu (blue) and 60~psu (brown) ocean salinity. Panel (f) shows the size dependence for the salinity forcing (equatorial minus polar salinity flux, two salinities are assumed) using dots and the temperature forcing (the freezing point difference under the equatorial and polar ice shell) using crosses.}}
    \label{fig:EOS-Hice-Heatflux}
  \end{figure*}

The system we consider here is sketched in Fig.~\ref{fig:EOS-Hice-Heatflux} -- a 56-km deep ocean covered by an ice shell that is about 20~km thick \footnote{The polar ice shell is likely to be thinner given that the tidal dissipation in the ice shell and the silicate core both amplifies over the poles \citep{Beuthe-2019:enceladuss}. In a situation where the equatorial ice shell is thinner, the conclusion here can still apply after reversing the sign of the circulation and heat transport.}. From atop, the ocean constantly loses heat through the ice shell due to heat conduction. The heat loss is faster over regions where ice is thin and over the poles where the ice surface temperature is low, as shown by the green curve in Fig.~\ref{fig:EOS-Hice-Heatflux}c. To balance the heat loss, the ice shell and the silicate core needs to produce heat. In this study, we will focus on the shell-heating scenario, and discuss the potential impacts of bottom heating in the end. Since the ice dissipation amplifies over the polar regions even if the ice is completely flat \citep{Beuthe-2018:enceladuss}, the equilibrium ice shell geometry will be poleward thinning. To study how ocean heat transport changes with the size of the moon, we assume the same ice geometry for all numerical models and analysis -- $H(\phi)=H_0+\Delta H P_2(\sin\phi)$ ($H_0=20$~km, $\Delta H=3$~km and $P_2$ is the 2nd order Legendre polynominal). This ice thickness profile is presented by the solid curve in Fig.~\ref{fig:EOS-Hice-Heatflux}b. The ocean is forced by heat and salinity fluxes from the ice shell. The ocean-ice heat exchange guarantees that the ocean temperature at the water-ice interface equals the local freezing point, which is lower under a thick ice shell because of the high pressure (see Eq.3 in the appendix). The ocean-ice salinity flux is prescribed to account for the brine rejection and fresh water input associated with the freezing/melting (shown by the dashed gray curve in Fig.~\ref{fig:EOS-Hice-Heatflux}b) required in order to keep the ice shell geometry in equilibrium assuming that ice dissipation can provide the required heating to sustain such freezing/melting \footnote{This is clearly not guaranteed in a real world, so when the required ice dissipation is too far off, the scenario should be considered as unphysical. The benefit of making such an assumption is that we cut off the positive feedback loop between ice freezing/melting and ocean circulation that can potentially make the climatology drift away. Details about the ocean general circulation model, the ice flow model, and the tidal dissipation model can be found in the appendix.}. Under these forcings, water over the poles will be warmer and fresher compared to the water at low latitudes. It should be noted that the setup here doesn't fully represent the thermodynamics near the ice-ocean interface and thus allows for some unphysical scenarios (such as ocean temperatures below the freezing point). A detailed investigation of thermodynamics near the ice-ocean interface is beyond the scope of this study but provides an interesting avenue for future research, which may reveal additional constraints to infer the ice-ocean interactions on icy satellites.

  Using parameters relevant for Enceladus, \cite{Kang-Mittal-Bire-et-al-2021:how} show that the circulation that arises from the surface heat and freshwater forcing can go either direction depending on the ocean salinity: in the low-salinity limit, temperature-induced density variation dominates, and the warm polar water would sink as sketched by the blue arrow in Fig.~\ref{fig:EOS-Hice-Heatflux}d because fresh water contracts upon warming (anomalous expansion); whilst in the high-salinity limit, the anomalous expansion is suppressed, and both salinity- and temperature-induced density gradients contribute to downwelling at low-latitudes, as sketched by the orange arrow in Fig.~\ref{fig:EOS-Hice-Heatflux}d . 

  When we consider icy satellites larger than Enceladus (most of the icy satellites of interest are), with everything else (such as the ice geometry, the vertical diffusivity and the roughness of the top and bottom boundaries) kept fixed, we qualitatively expect the following changes:
  \begin{itemize}
\item The thermal expansion coefficient near the freezing point will become more positive, and eventually anomalous expansion will be completely suppressed even if the ocean is relatively fresh. Shown in Fig.~\ref{fig:EOS-Hice-Heatflux}e are the dependence of thermal expansion coefficient on $a$ at the ocean-ice interface for two different salinities, 10~psu and 60~psu. Anomalous expansion doesn't occur in an ocean with 60~psu salinity regardless of the size of the satellite, yet it does occur in a 10-psu ocean, but only when the satellite's radius is smaller than 1500~km \footnote{20~km ice shell is assumed here.}, approximately the size of Europa. Therefore, on large icy satellites, we don't expect downwelling to ever occur over the poles.
\item The temperature difference under the ice shell between the equator and the pole $\Delta T$ will increase, as shown by the cross markers in Fig.~\ref{fig:EOS-Hice-Heatflux}f. The freezing point difference $\Delta T_f$ is proportional to the pressure difference $\Delta P$, which is equal to $\rho_ig\Delta H$ ($\rho_i$ is the ice density, $g$ is gravity and $\Delta H$ is the prescribe ice thickness difference). With $\rho_i$ and $\Delta H$ fixed, $\Delta T\sim \Delta T_f\propto \Delta P \propto g \propto a$.
\item The salinity forcing will weaken. The blue and brown dots in Fig.~\ref{fig:EOS-Hice-Heatflux}f show the salinity flux (mean salinity $S_0$ times the freezing rate $q$) difference between the equator and the poles for an ocean with mean salinity of 10~psu and 60~psu, respectively. The freezing rate $q$ is set to balance the divergence of ice flow. As derived in the appendix~A4, ice flow behaves like diffusion, and the flow divergence/convergence is proportional to $\Delta P$ divided by the distance square $a^2$, so $q \propto a^{-1}$. Combining this with the previous point, we anticipate temperature to play a more and more important role in driving ocean circulation, as the satellite's size increases.
\item The same density gradient will drive a stronger ocean circulation and heat transport as a result of the stronger gravity -- fixing the bulk density of the satellite, surface gravity $g$ should be proportional to the radius $a$. The stronger heat transport will then flatten the ice shell more efficiently.
  \end{itemize}

To illustrate the effect of the above changes, we use a conceptual model to predict how ocean heat transport changes with the size of the icy satellite, and how this affects the equilibrium ice thickness gradient. Similar to the Stommel box model \citep{Stommel-1961:thermohaline}, we consider two boxes, one representing the low latitudes (box1) and one representing the high latitudes (box2).

Driven by the density gradient between the two boxes $\Delta\rho$, we expect an overturning circulation to form: sinking in the dense box and rising in the buoyant box, with horizontal transport between the boxes near the top and bottom boundaries to close the loop. In lack of interior viscosity, the overturning cells in the ocean are forced to closely follow the axis of rotation to avoid the eastward/westward acceleration induced by conserving angular momentum \citep{Ashkenazy-Tziperman-2020:europas, Kang-Mittal-Bire-et-al-2021:how}. Only in the boundary layers, can angular momentum surfaces be readily crossed, thanks to the surface friction. This friction is represented by relaxing speeds to zero in the upper and lower boundaries. In equilibrium, the dominant balance of the zonal momentum equation in the boundary layer is
\begin{equation}
  \label{eq:Umom-boundary}
  -fV=-\gamma U/\delta z,
\end{equation}
where $\delta z$ is the depth of the boundary layer, and $\gamma=10^{-4}$~m/s is the surface friction rate. 

The meridional mass transport $\Psi\sim (\sqrt{2}\pi a \rho) V\delta z $ can then be related to the zonal flow $U$ through
\begin{equation}
  \label{eq:Psi-U}
  \Psi\sim \sqrt{2}\pi a \gamma \rho U/f.
\end{equation}
The amplitude of the zonal flow speed $U$ is governed by the thermal wind balance
\begin{equation}
  \label{eq:thermal-wind}
  U\sim \frac{\partial U}{\partial z} d=\frac{1}{f}\frac{\partial b}{\partial y} d\sim \frac{4gd\Delta \rho}{f\pi a \rho_0},
\end{equation}
where $d$ denotes the depth to which surface temperature and salinity anomalies can reach. This depth is set by the vertical diffusivity $\kappa_v$ via the advective-diffusive scaling, while limited by the depth of the ocean, $D$.
\begin{equation}
  \label{eq:adv-dif-psi}
  d= \min\{d_{\mathrm{diff}},D\}=\min\{\rho_0(2\pi a^2)\kappa_v/\Psi,D\},
\end{equation}
where $d_{\mathrm{diff}}$ is the advective-diffusive depth scale.
When $D$ is large enough, circulation and density gradient only penetrate to a depth $d$, such that the vertical heat advection $wT_z\sim w\Delta T/d$ ($w=\Psi/(2\pi a^2)$) is balanced by the vertical diffusion $\kappa_vT_{zz}\sim \kappa_v \Delta T/d^2$ \citep{Munk-1966:abyssal}.

Combining Eq.\ref{eq:Psi-U}, Eq.\ref{eq:thermal-wind} and Eq.\ref{eq:adv-dif-psi}, we obtain
\begin{equation}
  \label{eq:Psi-drho}
  \Psi\sim 
  \begin{dcases}
    A(a) \cdot\sqrt{|\Delta \rho/\rho_0|}\cdot \mathrm{sign}(\Delta \rho)=A_0a^{3/2}\cdot\sqrt{|\Delta \rho/\rho_0|}\cdot \mathrm{sign}(\Delta \rho),        & \text{if } d_{\mathrm{diff}}<D\\
    B(a) \cdot(\Delta \rho/\rho_0)=B_0a\cdot(\Delta \rho/\rho_0),          & \text{if } d_{\mathrm{diff}}\geq D
\end{dcases}
\end{equation}
The factor $A(a)=A_0a^{3/2}$ and $B(a)=B_0a$ characterizes the mobility of the ocean -- the mass exchange rate for a given equator-to-pole density contrast. $A_0$ and $B_0$ are given by
  \begin{eqnarray}
    A_0&=&\frac{4\pi \rho\sqrt{2^{3/2}G\rho_{\mathrm{bulk}}\kappa_v\gamma}}{\sqrt{3}f}    \label{eq:A0}\\
    B_0&=&\frac{16\sqrt{2}\pi\rho G \rho_{\mathrm{bulk}}D\gamma}{3f^2}, \label{eq:B0}
  \end{eqnarray}
  
  where $\rho_{\mathrm{bulk}}$ is the satellite's bulk density and $G$ is the gravitational constant. Since we fix all parameters except the moon's radius, $A_0,\ B_0$ are constants in this work. Assuming $\kappa_v=10^{-3}$~m$^2$/s and $\gamma=10^{-4}$~m/s, we get $A_0\sim 40$~kg/m$^{3/2}$/s and $B_0=2\times 10^7$~kg/m/s with Europa's parameters, and $A_0\sim 10$~kg/m$^{3/2}$/s and $B_0=10^6$~kg/m/s with Enceladus' parameters. The first case in Eq.\ref{eq:Psi-drho} corresponds to a scenario where circulation strength is limited by the vertical diffusivity $\kappa_v$ ($\kappa_v$-limit), whereas in the second case, the ocean depth becomes the limit ($D$-limit). Generally speaking, strong circulations tend to be in the $\kappa_v$-limit because strong upward motions ($\Psi\sim w(\pi a^2)$) require a large $\kappa_v$. It should also be noted that $\kappa_v$ itself may depend on $\Delta \rho$ in reality as more energy is needed to mix a more strongly stratified ocean. This is important because an energetically constrained diffusivity may change the scaling for the radius dependence of $\Psi$ and heat transport. We will use numerical model results to demonstrate this point later.  
  

  Both salinity and temperature gradients can induce density variations
  \begin{equation}
    \label{eq:rho-T-S}
    \Delta \rho/\rho_0=-\alpha_T\Delta T+\beta_S\Delta S,
  \end{equation}
  which in turn drives ocean circulation.
  The temperature contrast $\Delta T$ is determined by the pressure-induced freezing point shift from the north pole to the equator,
  \begin{equation}
    \label{eq:deltaT}
    \Delta T=b_0\Delta P=b_0\rho_i g \Delta H ,
  \end{equation}
  where $b_0=-7.61\times10^{-3}$~K/bar, $\rho_i=917$~kg/m$^3$ is the ice density, and $\Delta H\sim 3$~km is the difference in ice thickness between the equator and the north pole (see Fig.~\ref{fig:EOS-Hice-Heatflux}b). Since $g$ is proportional to $a$, $\Delta T$ increases linearly with $a$ as shown in Fig.~\ref{fig:EOS-Hice-Heatflux}f.

  The salinity contrast $\Delta S$ is not directly prescribed, however. Assuming the ice shell is in equilibrium, freezing/melting of the ice shell $q$ has to be able to compensate the smoothing effect of ice flow; this effect can be estimated using an ice flow model (Appendix~A4). In an ocean with nonzero salinity, ice freezing and melting will induce a salinity flux $S_0q$ into the ocean through brine rejection and freshwater production. When salinity distribution reaches an equilibrium, the ocean salinity transport, $\Psi\cdot \Delta S$, needs to balance the salinity forcing from atop, which gives
  \begin{equation}
    \label{eq:deltaS}
    \left|\Psi\right|\Delta S=\rho_0 S_0\Delta q \times (\pi a^2)
  \end{equation}
  Here, $\Delta q$ denotes the freezing rate difference between the equator and the pole in units of m/s. As shown by Fig.~\ref{fig:EOS-Hice-Heatflux}f, salinity flux per unit area $\rho_0S_0\Delta q$ decreases with radius $a$, meaning that salinity forcing will be less important for large moons.

  \begin{figure*}
    \centering \includegraphics[page=2,width=0.9\textwidth]{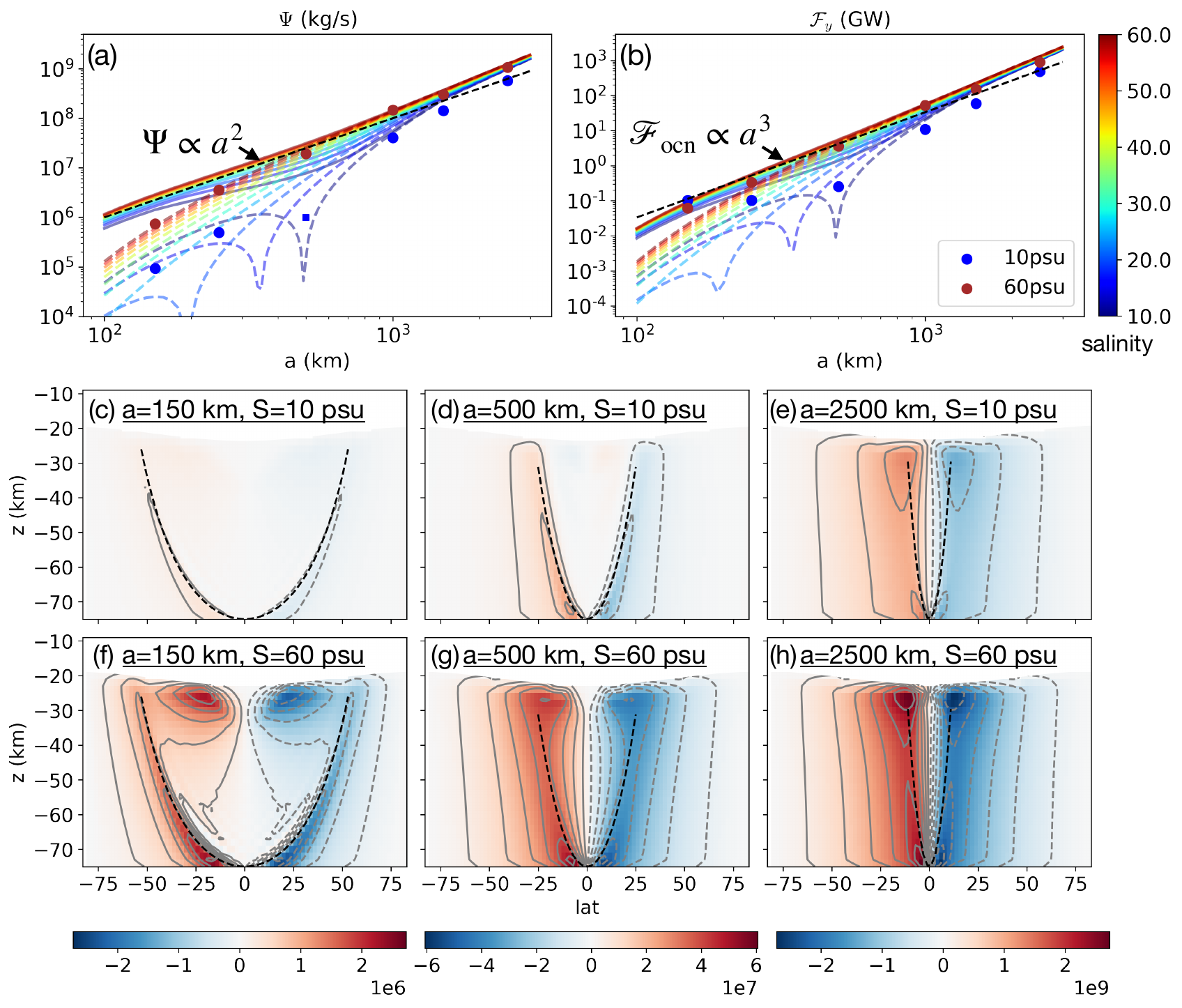}
    \caption{\small{Scaling laws for overturning streamfunction and meridional heat transport. Solid lines in panel (a,b) show the mass exchange rate $|\Psi|$ and equatorward heat transport $\mathcal{F}_y$ predicted by the conceptual model (Eq.\ref{eq:Psi-drho}-\ref{eq:deltaS}), and dashed lines show the scaling prediction given by Eq.~\ref{eq:Psi-Tlimit} and Eq.~\ref{eq:heat-transport} for various salinities. From blueish colors to reddish colors, salinity increases. The black dashed line in panel (a,b) denote the slope of $\Psi\propto a^2$ and $\mathcal{F}_{\mathrm{ocn}}\propto a^3$. Markers in panel (a) show the global mean overturning streamfunction amplitude $\overline{|\Psi|}$ diagnosed from 2D GCM simulations. When a reversed circulation is identified, the dot marker is replaced by a square marker. Markers in panel (b) show the maximum value of vertically-integrated equatorward heat convergence from the simulations. Blue markers are for 10~psu scenarios and red markers are for 60~psu scenarios. Panel (c-h) show the meridional streamfunction $\Psi$ for six representative cases in the GCM simulations. The upper row shows solutions with 10~psu salinity for $a=150,\ 500,\ 2500$~km and the lower row shows the same for 60~psu salinity  scenarios.}}
    \label{fig:scaling-circulation}
  \end{figure*}

  Since the purpose of this work is to understand how the satellite sizes and ocean salinities affect the circulation strength, we solve Eqs~(\ref{eq:Psi-drho})-(\ref{eq:deltaS}) for various $a$ and $S_0$ (other parameters can be found in appendix Table.~A1) and compare the conceptual model prediction against numerical model results. The following procedures are taken to obtain the prediction: we first assume that $\kappa_v$ is sufficiently high for circulation to reach the bottom, i.e. $D$-limited, and then examine whether the required $\kappa_v$ is greater than the assumed value $10^{-3}$~m$^2$/s; if so, we replace the $D$-limited solution with the $\kappa_v$-limited solution. For some scenarios, there are two stable solutions, similar to \citet{Marotzke-2000:abrupt}, in which case we only keep the solution with positive $\Psi$ (with sinking motions over the equator) because this is the solution that would be obtained if the initial salinity variation is set to zero, which is how the numerical simulations are initialized. The solution for the overturning circulation, $\Psi$, from the conceptual model is shown by lines in Fig.\ref{fig:scaling-circulation}a. Blueish colors here correspond to fresh oceans and reddish colors correspond to salty oceans. Most of the salty ocean solutions are $\kappa_v$-limited while most fresh ocean solutions are $D$-limited, especially those with intermediate satellite sizes (see appendix Fig.\ref{fig:Dlimit-kappalimit}a).


  Numerical model results for the two end-member ocean salinities ($S_0$=10~psu and 60~psu, with other parameters chosen to be the same as in the conceptual model) are shown by brown and blue dots in Fig.\ref{fig:scaling-circulation}a. As graphically presented in Fig.~\ref{fig:EOS-Hice-Heatflux}, these numerical experiments simulate the overturning circulation in the ocean forced by the salinity and heat forcing induced by a poleward-thinning ice shell. The upmost ocean grid is relaxed to the local freezing temperature which depends mostly on the pressure at the water-ice interface, and the salinity flux from the ice shell is prescribed such that it will balance the mass transport by ice flow (more details on the model setup can be found in the appendix~A). Under these forcings, the polar water is warmer and fresher than the equatorial water as shown by Fig.A1,A2 in the appendix. In most scenarios, sinking motions occur near the equator where water is colder and saltier, and thereby denser than water over the poles as shown by Fig.~\ref{fig:scaling-circulation}(c,e,f,g,h). However, due to the anomalous expansion of freshwater at low salinity and low pressure, cold water may become more buoyant than warm water, leading to a circulation in the opposite direction (see Fig.~\ref{fig:scaling-circulation}d). We note that these models are configured to be zonally symmetric to allow us to integrate them for tens of thousands of years until equilibrium is reached. Without the zonal dimension, eddies and eddy heat transport are not accounted for here, and their potential impacts are to be investigated in future studies. 

  As can be seen from Fig.\ref{fig:scaling-circulation}a, the mass exchange rate $\Psi$ varies roughly in proportion to $a^2$ (shown by the black dashed line) in both conceptual model and numerical simulations. This scaling law is expected when temperature becomes the dominant factor in determining the density as the moon's size increases. With $|\alpha \Delta T|\gg |\beta \Delta S|$, we can drop the salinity-induced density change (2nd term in Eq.~\ref{eq:rho-T-S}). Since $\Delta T\propto \Delta P\propto g\Delta H \propto a \Delta H$, the meridional mass transport follows from Eq.~\eqref{eq:Psi-drho}, \eqref{eq:rho-T-S} and \eqref{eq:deltaT} as
  \begin{equation}
    \label{eq:Psi-Tlimit}
    |\Psi|\sim 
  \begin{dcases}
    A_0a^{3/2}\sqrt{|-\alpha\Delta T|}\propto a^2\Delta H^{1/2},& \text{if } d_{\mathrm{diff}}<D\\
    B_0a|-\alpha\Delta T|\propto a^2\Delta H,     & \text{if } d_{\mathrm{diff}}\geq D
\end{dcases}
\end{equation}
The above results are shown by dashed curves in Fig.~\ref{fig:scaling-circulation}. There is a dip in each curve with low salinity due to the vanishing of $\alpha$. Overall, the simplified solutions seem to match the numerical results even better than the full conceptual model, probably due to a fortunate cancellation of errors. For icy moons with a larger size (Europa, Ganymede, Callisto, Titan, Triton, Pluto, etc.), the temperature effect will be certainly dominant, given that $\Delta T\propto a$ and $\Delta q\propto a^{-1}$. However, for smaller icy moons (such as Enceladus), the ocean circulation could be dominantly driven by salinity instead of temperature, and the above approximation could lead to an underestimation of the circulation strength as can be seen in Fig.~\ref{fig:scaling-circulation}.

The ocean circulations in the case with a relatively fresh ocean are generally weaker than the circulations in the salty cases because the thermal expansion coefficient $\alpha$ is small and the salinity flux is weaker when salinity is low (Fig.~\ref{fig:EOS-Hice-Heatflux}e). On top of this, $\Psi$ is further weakened or even reversed, when the moon's radius is around $500$~km. This behavior has to do with fresh water's anomalous expansion near the freezing point. With negative $\alpha$, the temperature-induced circulation acts against the salinity-induced one. Anomalous expansion maximizes when $|\Delta T|$ is large and $\alpha$ is very negative. This requires radius $a$ to be not too small because $|\Delta T|\propto a$, and to be not too large either because $\alpha$ becomes less negative with pressure, which varies proportional to $a$ as well (see blue dots in Fig.~\ref{fig:EOS-Hice-Heatflux}e). At 10psu salinity, an intermediate radius of $\sim 500$~km turns out to be the sweet spot here. When the radius is very close to 500~km, temperature-induced circulation dominates, driving downwelling over the mid-latitudes and upwelling over the equator, as can be seen in Fig.~\ref{fig:scaling-circulation}d. Unlike the results in \citet{Kang-Mittal-Bire-et-al-2021:how}, the circulation is only partially reversed because a nonlinear equation of state is used here and in the lower part of the ocean, anomalous expansion is suppressed by high pressure. When the radius is slightly off (e.g., the $150$~km, $250$~km and $1000$~km cases here), the circulation will just be weakened while sinking remains at the equator (Fig.~\ref{fig:scaling-circulation}c).

The relatively weak circulation make these scenarios with low salinity and small to  intermediate radius $D$-limited, unlike most other scenarios (see the appendix Fig.A3a for a comparison between $D$- and $\kappa_v$-limited solutions). According to Eq.\ref{eq:adv-dif-psi}, the main difference between the $\kappa_v$- and $D$-limited solutions lies in whether or not the density anomalies induced by surface forcing can penetrate all the way to the bottom. This is verified by numerical results. As shown by Fig.A1,A2, the temperature, salinity and density gradients are confined near the ice shell for the $\kappa_v$-limited scenarios (all salty scenarios plus those fresh scenarios with $a>2000$~km), whereas they are diffused through the entire ocean depth for the $D$-limited scenarios (fresh scenarios with $a<2000$~km). The default $\kappa_v$ and $\gamma$ yield very similar results under the $\kappa_v$- and $D$-limit, making it unclear which is more relevant (appendix Fig.A3). To distinguish them better, we did two sensitivity tests, one with 10 times larger friction rate $\gamma$ and the other with 10 times lower diffusivity $\kappa_v$. With these changes, $\kappa_v$ becomes more clearly the limiting factor (see Fig.A3c,e), so that $d \ll D$, as reflected by the shallower penetration of T, S, $\rho$ anomalies (see Fig.~A5,A6,A8,A9).

  Given the mass transport rate $\Psi$, we can estimate the meridional heat transport. Analogously to Eq~(\ref{eq:deltaS}), the equatorward heat transport $\mathcal{F}_{\mathrm{ocn}}$ between the two boxes can be written as
  \begin{equation}
    \label{eq:heat-transport}
    \mathcal{F}_{\mathrm{ocn}}=-C_p|\Psi|\Delta T=
    \begin{dcases}
    -C_pA_0a^{3/2}\sqrt{|-\alpha\Delta T|}\Delta T \propto a^3\Delta H^{3/2},& \text{if } d_{\mathrm{diff}}<D\\
    -C_pB_0a|-\alpha\Delta T|\Delta T \propto a^3\Delta H^2,     & \text{if } d_{\mathrm{diff}}\geq D
\end{dcases}
  \end{equation}
  To obtain the proportionality relations in the last step above, we use Eq.\ref{eq:deltaT} to substitute $\Delta T$ and the fact that $g\propto a$ with fixed bulk density. This predicted heat transport again well matches the results from the full conceptual model (curves in Fig.\ref{fig:scaling-circulation}b) and the numerical simulations (markers in Fig.\ref{fig:scaling-circulation}b), except the heat transport on small icy moons ($a<1000$~km) is underestimated because of the missing of salinity-driven circulation. Sensitivity tests presented in appendix Fig.A4 and Fig.A7 show consistent results.

  \section{The equilibrium equator-to-pole ice thickness gradient.}
  \label{sec:equilibrium-dH}

  \subsection{Scaling Theory}

\begin{figure*}
    \centering \includegraphics[page=3,width=\textwidth]{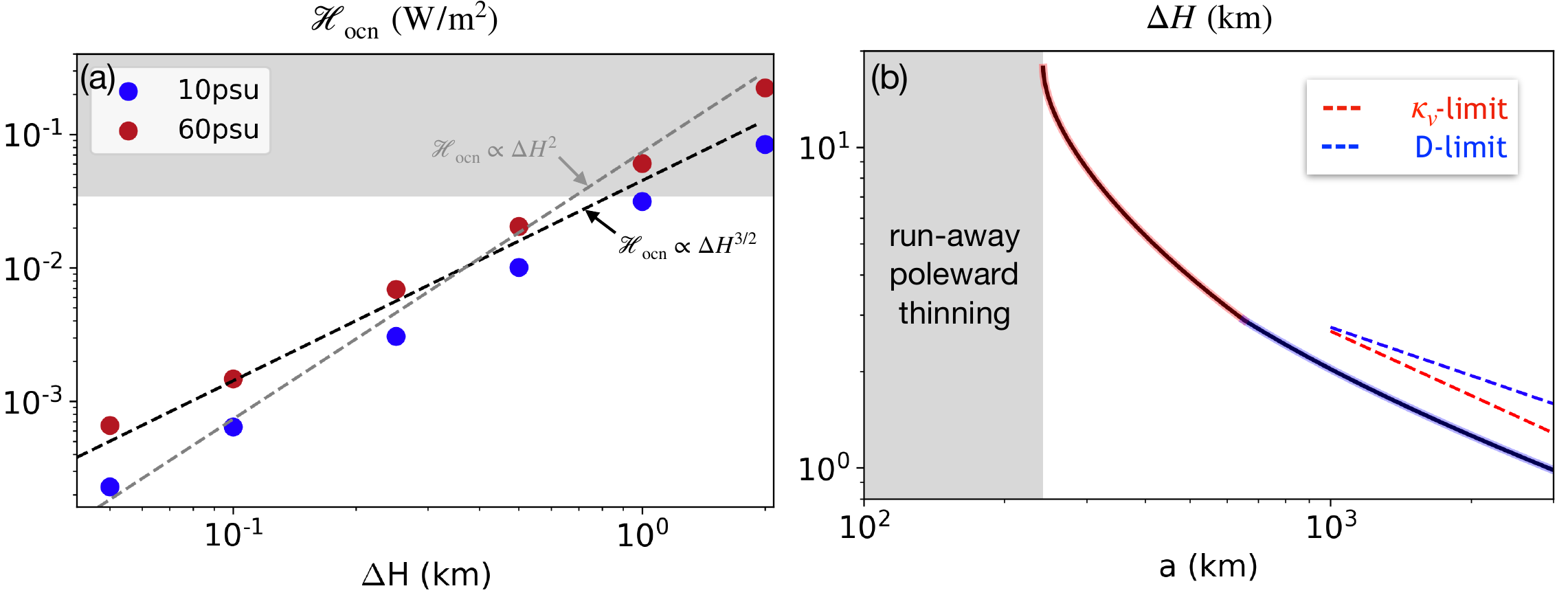}
    \caption{\small{Panel (a) shows the dependence of ocean meridional heat transport on equator-to-pole ice thickness gradient. The dashed line show the scaling law $\mathcal{F}_{\mathrm{ocn}}\propto \Delta H^2$, and markers show the vertically-integrated equatorward heat transport diagnosed from numerical models. Brown circles show results for high salinity scenarios (S=60~psu) and blue circles show results for low salinity scenarios (S=10~psu). Europa's size and orbital parameters are used in all cases. Gray shading is used to mark that the ocean heat transport is greater than what can be lost by conducting heat through the ice shell over an area of $\pi a^2$ ($a$ is the moon's radius). Solutions in this area hence are not consistent with an ice ice shell that is in equilibrium. Panel (b) shows the equilibrium ice thickness variation $\Delta H$ solved from Eq.~\eqref{eq:heat-balance}. The asymptotic scalings in the limits of large $\Delta H$ and small $\Delta H$ are shown by dashed lines. Reddish shading/lines indicate $\kappa_v$-limit and blueish shading/lines indicate $D$-limit. Gray shading denotes run-away poleward-thinning.}}
    \label{fig:heat-transport-europa-varydH}
  \end{figure*}

  In an equilibrium state, the meridional heat transport $\mathcal{F}_{\mathrm{ocn}}$ will be transmitted to the ice shell, distributed over about half a hemisphere with a surface area of $\pi a^2$, which yields a heat flux per unit area:
  \begin{widetext}
\begin{equation}
  \label{eq:Hocn}
  \mathcal{H}_{\mathrm{ocn}}=\left.\mathcal{F}_{\mathrm{ocn}}\right/(\pi a^2)=
\begin{dcases}
    \frac{A_0C_p(b_0\rho_i)^{3/2}|\alpha|^{1/2}}{\pi}\frac{g^{3/2}}{a^{1/2}}\Delta H^{3/2} \propto a \Delta H^{3/2},& \text{if } d_{\mathrm{diff}}<D\\
    \frac{B_0C_p (b_0\rho_i)^2 |\alpha|}{\pi}  \frac{g^2}{a} \Delta H^2\propto a \Delta H^2,     & \text{if } d_{\mathrm{diff}}\geq D
\end{dcases}
\end{equation}
\end{widetext}
The mean water-ice heat exchange rate $\mathcal{H}_{\mathrm{ocn}}$ is proportional to $a\Delta H^{3/2}$ or $a\Delta H^2$ depending on the magnitude of $\kappa_v$. We try to test this scaling using a set of numerical experiments forced by various $\Delta H$ from 100~m to 2~km (using $S_0=60$~psu, and other parameters relevant for Europa; see appendix Table.\ref{tab:parameters} for details). The default parameters should put the model near the boundary between the D-limited and kappa-limited regimes, and therefore, the numerical results fall between the D-limited scaling, $\mathcal{H}_{\mathrm{ocn}}\propto \Delta H^{2}$, and the $\kappa_v$-limited scaling, $\mathcal{H}_{\mathrm{ocn}}\propto \Delta H^{3/2}$, shown by Fig.~\ref{fig:heat-transport-europa-varydH}a. When $\gamma$ is chosen to be ten times larger, the $\kappa_v$-limited scaling governs instead (see appendix Fig.A10), because the circulation becomes much stronger and strong temperature gradients are confined to the upper ocean (see appendix Fig.A4). 

When equilibrium is reached, the ice shell needs to be in a heat balance everywhere. This requires
\begin{equation}
  \label{eq:heat-budget}
  \mathcal{H}_{\mathrm{ice}}+\mathcal{H}_{\mathrm{latent}}+\mathcal{H}_{\mathrm{ocn}}=\mathcal{H}_{\mathrm{cond}},
\end{equation}
where $\mathcal{H}_{\mathrm{ice}}$ denotes the ice dissipation, $\mathcal{H}_{\mathrm{latent}}$ denotes the latent heat release and $\mathcal{H}_{\mathrm{cond}}$ denotes the conductive heat loss.
The latent heat release $\mathcal{H}_{\mathrm{latent}}=\rho_iL_fq$ tends to be small compared to the other terms, as can be seen from Fig.~\ref{fig:EOS-Hice-Heatflux}c, except when the moon size is very small and the ice thickness variation is very large. For simplicity, we drop $\mathcal{H}_{\mathrm{latent}}$. The remaining terms can all be written as a function of the ice topography $H$. $\mathcal{H}_{\mathrm{ice}}$ amplifies over regions with thinner ice due to the rheology feedback \citep{Beuthe-2019:enceladuss, Kang-Flierl-2020:spontaneous} following $\mathcal{H}_{\mathrm{ice}}(\phi)=\mathcal{H}_{\mathrm{ice0}}(\phi)\times(H_0/H(\phi))^2$, where $H_0$ is the mean ice thickness, $\mathcal{H}_{\mathrm{ice0}}(\phi)$ is the ice dissipation rate in a flat ice shell as a function of latitude $\phi$. The conductive heat loss $\mathcal{H}_{\mathrm{cond}}$ is inversely proportional to the local ice thickness $\mathcal{H}_{\mathrm{cond}}(\phi)=\mathcal{H}_{\mathrm{cond0}}\times(H_0/H(\phi))$. Since the ice dissipation over the polar regions is roughly twice as strong as that over the equator in absence of ice topography \citep{Beuthe-2018:enceladuss}, we choose $\mathcal{H}_{\mathrm{ice0}}(\mathrm{pole})=1.25\overline{\mathcal{H}_{\mathrm{ice0}}}$ for the polar box and $\mathcal{H}_{\mathrm{ice0}}(\mathrm{eq})=0.75\overline{\mathcal{H}_{\mathrm{ice0}}}$ for the equatorial box, where $\overline{(\cdot)}$ denotes the global mean. To guarantee global heat budget balance, $\overline{\mathcal{H}_{\mathrm{ice0}}}$ should be equal to $\mathcal{H}_{\mathrm{cond0}}\overset{\Delta}{=} \mathcal{H}$. Now, we consider the equator-to-pole difference of the heat budget terms for an ice shell that has a mean thickness of $H_0$ and an equator-to-pole thickness difference of $\Delta H$ (equatorial ice shell is thicker), which yields
\begin{widetext}
\begin{equation}
  \label{eq:heat-balance}
  2\mathcal{H}_{\mathrm{ocn}}=\Delta \mathcal{H}_{\mathrm{ice}}-\Delta \mathcal{H}_{\mathrm{cond}}\approx 1.25\mathcal{H}\left(1-\frac{\Delta H}{2H_0}\right)^{-2}-0.75\mathcal{H}\left(1+\frac{\Delta H}{2H_0}\right)^{-2}-\left[\mathcal{H}\left(1-\frac{\Delta H}{2H_0}\right)^{-1}-\mathcal{H}\left(1+\frac{\Delta H}{2H_0}\right)^{-1}\right].
\end{equation}
\end{widetext}
From Eqs.~(\ref{eq:heat-balance}) and (\ref{eq:Hocn}), one can solve for $\Delta H$ numerically, and the results are presented in Fig.~\ref{fig:heat-transport-europa-varydH}b. Red shading marks the $\kappa_v$-limit regime and blue shading marks the $D$-limit regime.

Further simplification can be made to Eq.~\eqref{eq:heat-balance} if the ice thickness variation is small ($\Delta H/H_0\ll 1$). Expanding for small $\Delta H/H_0$ and keeping only the lowest order terms, we get
\begin{equation}
  \label{eq:heat-balance_small_dH}
  2\mathcal{H}_{\mathrm{ocn}} \approx \frac{1}{2}\mathcal{H}
\end{equation}
From Eqs. (\ref{eq:heat-balance_small_dH}) and (\ref{eq:Hocn}) we finally get
\begin{equation}
   \label{eq:limit-small-dH}
   \Delta H\approx
   \begin{dcases}
    \left(\frac{\chi_\kappa }{4\kappa_v\gamma|\alpha|}\right)^{1/3}\left(\frac{f}{a}\right)^{2/3},& \text{if } d_{\mathrm{diff}}<D\\
    \left(\frac{\chi_D}{2\gamma D|\alpha|}\right)^{1/2}\frac{f}{a^{1/2}},     & \text{if } d_{\mathrm{diff}}\geq D
\end{dcases}
\end{equation}

where $\chi_\kappa=\left.(81\mathcal{H}^2)\right/[8192\sqrt{2}\pi^3\rho C_p^2b_0^3(G\rho\rho_{\mathrm{bulk}})^4]$ and $\chi_D=\left.(27\mathcal{H})\right/[1048\sqrt{2}\pi^2 C_pb_0^2(G\rho\rho_{\mathrm{bulk}})^3]$ are constants, in which we have already substituted the definitions of $A_0$ and $B_0$ (Eq.~\ref{eq:A0} and \ref{eq:B0}), and the surface gravity $g=4\pi G\rho_{\mathrm{bulk}}a/3$ (where $\rho_{\mathrm{bulk}}=2500$~kg/m$^3$ is the bulk density). 


From Eq.~(\ref{eq:limit-small-dH}), it can be seen that, when $\Delta H\ll H_0$, the equilibrium ice shell thickness variation $\Delta H$ should decrease with the moon's size and increases with the rotation frequency, following $\Delta H\propto (f/a)^{2/3}$ or $f/a^{1/2}$ depending on the dynamic regime of the ocean. The asymptotic scalings (shown by blue and red dashed lines in Fig.~\ref{fig:heat-transport-europa-varydH}b) provide a useful approximation to the full solution of Eq.~\eqref{eq:heat-balance} for relatively small to moderate $\Delta H$. For larger $\Delta H$, the sensitivity of $\Delta H$ on $a$ increases, and eventually, a run-away poleward-thinning happens when $a\approx 250$~km (masked by gray shading), due to the strengthening of the ice-rheology feedback.
Besides the dependence on $a$ and $f$, a higher ocean salinity (leading to larger $\alpha$ and stronger salinity-driven circulation), a stronger turbulent diffusivity, and (provided sufficient turbulent mixing) a deeper ocean can also reduce $\Delta H$.

Compared to Enceladus, Europa is $6$ times larger in size, is rotating $3$ times slower, and its ocean is likely saltier \citep{Zolotov-2007:oceanic, zolotov2014can, Glein-Postberg-Vance-2018:geochemistry, Kang-Mittal-Bire-et-al-2021:how, Hand-Chyba-2007:empirical} and deeper \citep{Hand-Chyba-2007:empirical} -- all these differences suggest that Europa's ice shell may be much flatter than Enceladus'. Assuming $|\alpha|\sim 10^{-5}$-$10^{-4}$/K (Fig.~\ref{fig:EOS-Hice-Heatflux}e), $\kappa_v=10^{-3}$~m$^2$/s, $\gamma=10^{-4}$~m/s, Eq.\ref{eq:heat-balance} yields run-away solutions ($\Delta H>H_0$) for Enceladus \footnote{In reality, $\Delta H$ should be smaller if we account for 1) the salinity-driven circulation, which could be crucial for small icy satellites, 2) $\mathcal{H}_{\mathrm{latent}}$ in the heat budget (Eq.\ref{eq:heat-budget}) and 3) the nonlinear effects that make $\mathcal{H}_{\mathrm{ocn}}$ more sensitive than predicted by the scaling law (Fig.~\ref{fig:heat-transport-europa-varydH}a).}, in line with the strong ice topography in observations \citep{Iess-Stevenson-Parisi-et-al-2014:gravity, Beuthe-Rivoldini-Trinh-2016:enceladuss, Tajeddine-Soderlund-Thomas-et-al-2017:true, Cadek-Soucek-Behounkova-et-al-2019:long, Hemingway-Mittal-2019:enceladuss}. In contrast, using Europa parameters,  $\Delta H$ is estimated to be only 1-3~km! This is consistent with the varying-$\Delta H$ experiments shown in Fig.~\ref{fig:heat-transport-europa-varydH}a, and it also roughly matches a constraint based on limb profile measurements \citep{Nimmo-Thomas-Pappalardo-et-al-2007:global}. Finally, it should be noted that the vertical diffusivity and the friction rate also have a strong impact on $\Delta H$. Since these parameters are poorly constrained for icy moon oceans, the quantitative results here should be interpreted with care.

\subsection{Numerical Results for Enceladus and Europa}
 \label{sec:evolution}
 
 To demonstrate the potential impacts of the size of the icy satellite on its equilibrium ice shell geometry, we integrate an ice evolution model forward using Enceladus' and Europa's parameters, respectively. The model is modified based on \citet{Kang-Flierl-2020:spontaneous}. It calculates the melting induced by the tidal heating $\mathcal{H}_{\mathrm{ice}}$ (given by Eq.\ref{eq:H-tide} in the appendix), the down-gradient ice flow $\mathcal{Q}$ (given by Eq.\ref{eq:ice-flow} in the appendix), the heat loss to space by conduction $\mathcal{H}_{\mathrm{cond}}$ (given by Eq.\ref{eq:H-cond} in the appendix), and heat transmitted upward by the ocean $\mathcal{H}_{\mathrm{ocn}}$, and evolves the ice thickness $H$ over time. The total thickness tendency can be symbolically expressed as 
\begin{equation}
  \label{eq:ice-evolution}
  \frac{dH}{dt}=\frac{\mathcal{H}_{\mathrm{cond}}(H)-\mathcal{H}_{\mathrm{ice}}(H)-\mathcal{H}_{\mathrm{ocn}}(H)}{L_f\rho_i } + \frac{1}{a\sin\phi}\partial_\phi \left(\sin\phi \mathcal{Q}(H)\right),
\end{equation}
where $L_f$ and $\rho_i$ are the latent heat of freezing and density for ice, $a$ is the moon's radius and $\phi$ denotes latitude. When the ice shell is thinner than 3~km, we assume that the ice shell will crack open under the tidal stress, and the resultant geysers will carry away large amounts of heat, preventing further melting. In the evolution model, we overwrite the thickness tendency with zero when $H<3$~km, to implicitly represent this additional heat loss.

The ocean-ice heat exchange $\mathcal{H}_{\mathrm{ocn}}$ is a new component that doesn't exist in \citet{Kang-Flierl-2020:spontaneous}. Inspired by the conceptual model in Eq. (\ref{eq:Hocn}), $\mathcal{H}_{\mathrm{ocn}}$ is parameterized as
\begin{eqnarray}
  \label{eq:H-ocn}
  \mathcal{H}_{\mathrm{ocn}}&=&\mathrm{MIN}\{(2A_0C_p(b_0\rho_i)^{3/2}|\alpha|^{1/2}/\pi)\frac{g^{3/2}}{a^{1/2}}(2|H'|)^{3/2}\nonumber\\
  &~&,\ (2B_0C_p (b_0\rho_i)^2 |\alpha|/\pi)  \frac{g^2}{a} (2|H'|)^2\}\cdot \mathrm{sign}(H'),
\end{eqnarray}
where $H'$ is the deviation from the prescribed global mean ice thickness $H_0=20$~km and $\mathrm{MIN}\{\}$ selects whichever parameterization yields a smaller global standard deviation, which is equivalent to selecting the depth scale as $d=\mathrm{MIN}\{d_{\mathrm{diff}},D\}$ as done in Eq.~\eqref{eq:adv-dif-psi}. A factor of 2 is multiplied to H' because $\Delta H$ in Eq. (\ref{eq:Hocn}) refers to the equator to pole thickness difference, which is twice as large as the equator/pole's deviation from the mean, and we multiply an additional factor of 2 to the ocean heat transport based on the assumption that the salinity-driven heat transport is roughly comparable with the temperature-driven one.

\begin{figure*}
    \centering \includegraphics[page=4,width=\textwidth]{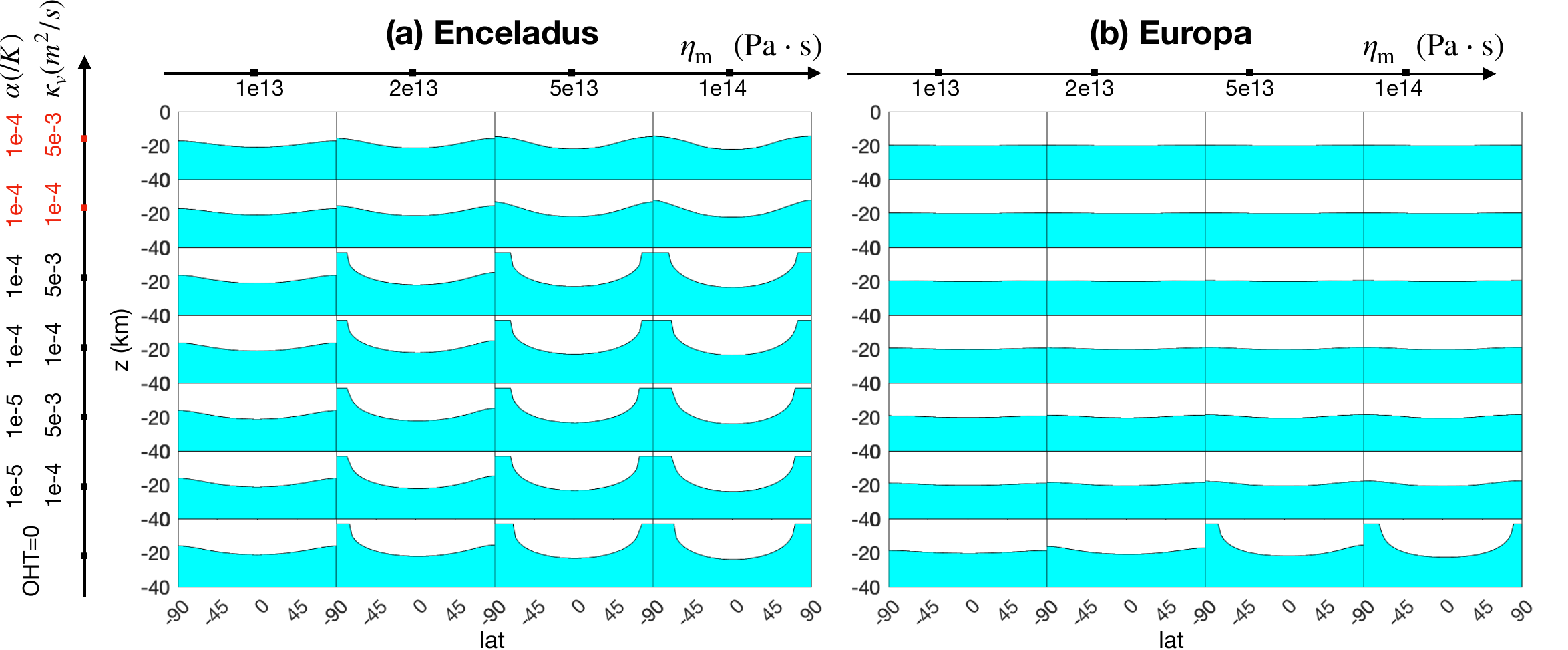}
    \caption{\small{Equilibrium ice shell geometries on Enceladus (panel a) and Europa (panel b) predicted by an ice evolution model (Eq.~\ref{eq:ice-evolution}) with parameterized ocean heat transport (Eq.~\ref{eq:H-ocn}). Blue color masks the ocean and white color masks the ice. 24 scenarios are considered for each moon, to account for the uncertainties associated with the ice shell rheology and the efficiency of ocean heat transport. The values for the three key parameters, $\eta_m$, $|\alpha|$ and $\kappa_v$, are shown on the left and upper sides. The top two rows have ocean heat transport amplified by another factor of 10 to represent the potential effect of eddies and other unforeseeable factors. The bottom row assumes zero ocean heat transport.}}
    \label{fig:ice-evolution}
  \end{figure*}

To account for the uncertainties associated with the ice shell rheology and the efficiency of ocean heat transport, we explore a range of ice viscosities $\eta_m$ \footnote{Notice that a smaller $\eta_m$ leads to a stronger ice flow, thus requiring more freezing and melting.  As a result a small $\eta_m$ also affects the latent heating and the salinity-driven circulation, neither of which was included when deriving Eq.~\ref{eq:limit-small-dH} and Eq.\ref{eq:H-ocn}. Our model is therefore most adequate in the limit of large ice viscosity.}, $|\alpha|$ and $\kappa_v$ for Enceladus and Europa. The equilibrium ice shell geometries are shown in Fig.~\ref{fig:ice-evolution}. The bottom row assumes no ocean heat transport, and significant ice thickness variations develop on both Enceladus and Europa. When the ice viscosity is not too low ($\eta_m>10^{-13}$~Pa$\cdot$s), ice almost completely vanishes over one or both poles due to the ice-rheology feedback. However, with ocean heat transport, the equilibrium ice geometry is largely flattened especially for Europa thanks to its large size and slower rotation rate.  Since Europa's ocean is likely saltier than 50~psu \citep{Hand-Chyba-2007:empirical}, $\alpha$ should be closer to the upper bound, leading us to the conjecture that ice thickness variations on Europa are likely less than 2~km, if the vertical diffusivity on Europa is not significantly lower than $10^{-4}$~m$^2$/s. When Enceladus' parameters are used instead, the ocean heat transport's impact on ice shell geometry is more limited. However, if heat transport is enhanced by a factor of 10 (perhaps as a result of eddy heat transport, which is missing in our zonally symmetric model)
, the equilibrium ice geometry will again be effectively flattened (row-1 and 2). Among the 28 scenarios considered for Enceladus, five develop the hemispheric asymmetry seen in observations \citep{Iess-Stevenson-Parisi-et-al-2014:gravity, Hemingway-Mittal-2019:enceladuss} -- one pole reaches the 3~km limit, while the other is still around 10~km thick. For each individual moon, the equilibrium ice shell tends to be flatter with smaller $\eta_m$ and higher $|\alpha|$ and $\kappa_v$.

\section{Concluding remarks.}

This work studies how the size of icy worlds affects the efficiency of ocean heat transport and thereby the ice geometry. Using an ocean box model, scaling laws for the ocean overturning circulation $\Psi$ and ocean heat transport $\mathcal{F}_{\mathrm{ocn}}$ are found. These scaling laws are verified by 2D general circulation model simulations for various planetary radii $a$ and associated ice thickness variations $\Delta H$. The results show that heat is converged to thick-ice regions more efficiently on larger icy satellites, ultimately due to their higher gravity. This implies a more effective "ice pump" mechanism \citep{Lewis-Perkin-1986:ice} on larger satellites and thus their ice shells are expected to be flatter. 

We apply this result to Enceladus and Europa which are known to contain a global subsurface ocean \citep{Postberg-Kempf-Schmidt-et-al-2009:sodium, Thomas-Tajeddine-Tiscareno-et-al-2016:enceladus, Carr-Belton-Chapman-et-al-1998:evidence, Kivelson-Khurana-Russell-et-al-2000:galileo, Hand-Chyba-2007:empirical}. Enceladus' ice shell exhibits strong equator-to-pole ice thickness variations and a significant hemispheric asymmetry \citep{Iess-Stevenson-Parisi-et-al-2014:gravity, Beuthe-Rivoldini-Trinh-2016:enceladuss, Tajeddine-Soderlund-Thomas-et-al-2017:true, Cadek-Soucek-Behounkova-et-al-2019:long, Hemingway-Mittal-2019:enceladuss}, whereas Europa's ice shell appears to be relatively flat \citep{Nimmo-Thomas-Pappalardo-et-al-2007:global}. Their difference may be explained by the mechanism proposed here. Based on the scaling laws for ocean heat transport, we parameterize the ocean-ice heat exchange for an arbitrary ice shell geometry and use this parameterization in the ice shell evolution model by \citet{Kang-Flierl-2020:spontaneous}. Integrating this model using Europa' parameters indeed leads to an equilibrium ice shell geometry that is much flatter than that obtained using Enceladus' parameters. 

In this work, we assumed that all heat is produced in the ice shell. If part of heating is generated in the silicate and core, the reduced heat production in the ice shell will likely weaken the ice shell thickness variations unless the heat distribution from the core is also strongly polar-amplified. Also, with heating from the silicate core, the ocean's stratification will change: a fresh ocean on a small moon with negative $\alpha$ will become more stratified, whereas a salty ocean on a large moon with positive $\alpha$ will become less stratified and even globally convective \citep{Melosh-Ekholm-Showman-et-al-2004:temperature, Zeng-Jansen-2021:ocean}. In the latter case, a 3D high-resolution numerical model is required to resolve the convection and associated heat transport. However, assuming fixed ice geometry and a positive thermal expansion coefficient, we expect the effect of bottom heating to become weaker as the satellite size increases, because the vertical temperature gradient induced by bottom heating will decrease \footnote{According to \citet{Gastine-Wicht-Aubert-2016:scaling}, Nusselt number is proportional to Rayleigh number to the power of $1.5$. Solving the vertical temperature gradient $\Delta T$ assuming fixed vertical heat flux yields $\Delta T \propto g^{-3/5}$.} whilst the temperature difference induced by ice topography will increase with satellite size.

Another simplification we made in this work is to ignore the variability in the zonal (east and west) direction. This zonally symmetric framework is necessary in order to keep the computational cost of integrating an ocean circulation model for tens of thousands of years manageable. However, baroclinic eddies may play a key role in transporting heat meridionally. In the context of Earth's atmosphere, the poleward heat transport driven by the inhomogeneous solar heating is found to be significantly stronger in a 3D setup than a 2D setup, due to the contribution by eddies \citep{Schneider-2006:general}. Besides transporting heat, barolinic eddies also redistribute zonal momentum, which gives rise to the mid-latitude jet on Earth and the jets on Jupiter \citep{Liu-Schneider-2010:mechanisms, Kaspi-Flierl-2007:formation}. Neither of these processes is captured in the 2D configuration here.

Despite these shortcomings, the qualitative result that ocean heat transport is more efficient at limiting ice-shell thickness variations on large satellites is likely to be robust. Future missions that can improve our estimates of the ice thickness distribution on icy moons will provide evidence to support of reject this hypothesis.

\begin{acknowledgments}
  This work is carried out in the Department of Earth, Atmospheric and Planetary Science (EAPS) in MIT. WK acknowledges support as a Lorenz-Houghton Fellow by endowed funds in EAPS. 
\end{acknowledgments}

%



\software{MITgcm \citep{MITgcm-group-2010:mitgcm}}



\appendix
\setcounter{figure}{0}
\renewcommand{\thefigure}{A\arabic{figure}}
\setcounter{table}{0}
\renewcommand{\thetable}{A\arabic{table}}
\section{A description of the General Circulation Model.}

  
  Our simulations are carried out using the Massachusetts Institute of Technology OGCM \citep[MITgcm][]{MITgcm-group-2010:mitgcm, Marshall-Adcroft-Hill-et-al-1997:finite} configured for application to icy moons. Our purpose is to 1) simulate the large-scale circulation and tracer transport driven by under-ice salinity gradients induced by patterns of freezing and melting and under-ice temperature gradients due to the pressure-dependence of the freezing point of water, 2) diagnose the meridional heat transport by the ocean and 3) examine the heat transport against the scaling laws.
  
  In our calculations the ice shell freezing/melting rate is derived from a model of ice flow (described below), based on observational inferences of ice shell thickness, prescribed and held constant: it is not allowed to respond to the heat/salinity exchange with the ocean underneath. To enable us to integrate our ocean model out to equilibrium on a 10,000 year timescale and to explore a wide range of parameters, we employ a zonally-symmetric configuration at relatively coarse resolution, and parameterize the effect of processes such as baroclininc instability, convection and small-scale turbulence, which cannot be resolved. Each experiment is initialized from rest and a constant salinity distribution. The initial potential temperature at each latitude is set to be equal to the freezing point at the water-ice interface. The simulations are then integrated for 10,000 years. By the end of 10,000 years of integration thermal equilibrium has been reached.
  
  The model integrates the non-hydrostatic primitive equations for an incompressible fluid in height coordinates, including a full treatment of the Coriolis force in a deep fluid, as described in \citet{MITgcm-group-2010:mitgcm, Marshall-Adcroft-Hill-et-al-1997:finite}. Such terms are typically neglected when simulating Earth's ocean because the ratio between the fluid depth and horizontal scale is small. Instead, when the moon size is order hundreds of kilometers like Enceladus, the aspect ratio is order $0.1$ and so not negligibly small. The size of each grid cell shrinks with depth due to spherical geometry and is accounted for by using the ``deepAtmosphere'' option of MITgcm. Gravity also varies with depth, which is accounted for by using the following profile:
  \begin{equation}
    \label{eq:g-z}
    g(z)=\frac{4\pi G\left[\rho_{\mathrm{core}}(a-D_0-H_0)^3+\rho_{\mathrm{out}}((a-z)^3-(a-D_0-H_0)^3)\right]}{3(a-z)^2}.
  \end{equation}
  In the above equation, $G=6.67\times10^{-11}$~N$\cdot$m$^2$/kg$^2$ is the gravitational constant, $\rho_{\mathrm{core}}=2500$~kg/m$^3$ is the assumed core density and $\rho_{\mathrm{out}}=1000$~kg/m$^3$ is the density of the ocean/ice layer. $D_0$ and $H_0$ are the globally averaged thickness of the ocean and ice, respectively.
  
  Since it takes several tens of thousands of years for our solutions to reach equilibrium, we employ a moderate resolution of $2$~degree ($8.7$~km) and run the model in a 2D, zonal-average configuration whilst retaining full treatment of Coriolis terms. By doing so, the zonal variations are omitted (the effects of 3D dynamics are to be explored in future studies). In the vertical direction, the $60$~km ocean-ice layer is separated into $30$ layers, each of which is $2$~km deep. The ocean is encased by an ice shell with meridionally-varying thickness, assuming hydrostacy (i.e., ice is floating freely on the water). The ice thickness is set to be
  \begin{equation}
    \label{eq:Hice}
    H(\phi)=H_0-H_2P_2(\sin\phi),
  \end{equation}
  where $H_0$ is the mean ice thickness, $P_2$ is the 2nd order Legendre polynomial, and $H_2$ is the amplitude of the ice thickness variation. $\phi$ denotes latitude. The thickness profile is shown by a solid curve in Fig.1b of the main text. We employ partial cells to better represent the ice topography: water is allowed to occupy a fraction of the height of a whole cell with an increment of 10\%. Interactions between the ice shell and the ocean are taken care of by a modified version of the MITgcm's ``shelfice'' module \citep{Losch-2008:modeling}, as described below.

  \subsection{Parameterization of subgridscale processes}
  Key processes that are not explicitly resolved in our model are diapycnal mixing, convection and baroclinic instability. Here we review the parameterizations and mixing schemes used in our model to parameterize them. 
  
  \underline{Mixing of tracers and momentum}
  
  To account for the mixing of heat and salinity by unresolved turbulence, in our calculations, we set the explicit vertical diffusivity to $0.001$~m$^2$/s, following \citet{Kang-Mittal-Bire-et-al-2021:how}. This is roughly 4 orders of magnitude greater than molecular diffusivity, but broadly consistent with dissipation rates from libration-generated turbulence suggested by \citet{Rekier-Trinh-Triana-et-al-2019:internal} for Enceladus, using the scaling in \citet{Wunsch-Ferrari-2004:vertical}. 
  
  The horizontal viscosity is set to $\frac{a}{150~\mathrm{km}}$~m$^2$/s ($a$ is the radius of the moon) to control grid-scale noise. In addition, to damp numerical noise induced by our use of stair-like ice topography, we employ a bi-harmonic hyperviscosity of $\frac{a}{150~\mathrm{km}}\times 10^8$~m$^4$/s. The relatively high viscosity here may suppress some dynamic modes in the system, which may otherwise make the circulation and heat transport even stronger.
 Despite use of these viscous and smoothing terms, the dominant balance in the meridional momentum equation is between the Coriolis force and the pressure gradient force and so zonal currents on the large-scale remain in thermal wind balance, especially in the interior of the ocean. 
 
  \underline{Convection}
  
  Due to the coarse resolution of our model, convection cannot be resolved and must be parameterised. Following \citet{Kang-Mittal-Bire-et-al-2021:how}, we set the diffusivity to a much larger value in convectively unstable regions, to represent the vertical mixing associated with convective overturns. This convective diffusivity $\kappa_{\mathrm{conv}}$ is set to increase from $1$~m$^2$/s to $30$~m$^2$/s as gravity and convection strengthens with the satellite radius. Similar approaches are widely used to parameterize convection in coarse resolution ocean models (see, e.g. \citet{Klinger-Marshall-Send-1996:representation}) and belong to a family of convective adjustment schemes. Our results turn out to be insensitive to $\kappa_{\mathrm{conv}}$, as long as the convective timescale $D^2/\kappa_{\mathrm{conv}}< 1$~yr is much shorter than the advective time scale $M_{\mathrm{half}}/\Psi\approx 10^2$-$10^3$~yrs ($M_{\mathrm{half}}$ is half of the total mass of the ocean and $\Psi$ is the maximum meridional streamfunction in $kg/s$), and the convective instability is efficiently removed by the efficient diffusion in the unstable regions.

  \subsection{Equation of state and the freezing point of water}
  To make the dynamics as realistic as possible, we adopt the ``MDJWF'' equation of state \citep[EOS][]{McDougall-Jackett-Wright-et-al-2003:accurate} to determine how density depends on temperature, salinity and pressure. As demonstrated in Fig.~1e in the main text, the thermal expansion coefficient $\alpha$ at the freezing point is negative at the ice-ocean interface when the moon size is small (low pressure) and  the ocean is fresh. This anomalous expansion can suppress the convection driven by bottom heating \citep{Melosh-Ekholm-Showman-et-al-2004:temperature, Zeng-Jansen-2021:ocean} 
  and can alter the direction of ocean circulation \citep{Kang-Mittal-Bire-et-al-2021:how}.
  
  The freezing point of water $T_f$ is assumed to depend on local pressure $P$ and salinity $S$ as follows,
  \begin{equation}
    \label{eq:freezing-point}
    T_f(S,P)=c_0+b_0P+a_0S,
  \end{equation}
where $a_0=-0.0575$~K/psu, $b_0=-7.61\times10^{-4}$~K/dbar and $c_0=0.0901$~degC. The pressure $P$ can be calculated using hydrostatic balance $P=\rho_igH$ ($\rho_i=917$~kg/m$^3$ is the density of the ice and $H$ is the ice thickness).

  \subsection{Boundary conditions}
  \label{sec:boundary-conditions}
  
  Our ocean model is forced by heat and salinity fluxes from the ice shell at the top as well as heat fluxes coming from below.
  
  \underline{Diffusion of heat through the ice}
  
  Heat loss to space by heat conduction through the ice $\mathcal{H}_{\mathrm{cond}}$ is represented using a 1D vertical heat conduction model,
\begin{equation}
  \mathcal{H}_{\mathrm{cond}}=\frac{\kappa_{0}}{H} \ln \left(\frac{T_{f}}{T_{s}}\right),
  \label{eq:H-cond}
  \end{equation}
  where $H$ is the thickness of ice (solid curve in Fig.1b of the main text), $\kappa_0=651$~W/m is the heat conductivity of ice, $T_s$ is the surface temperature and $T_f$ denotes the local freezing point (Eq.~\ref{eq:freezing-point}) which equals to the ice temperature at the water-ice interface.
  We approximate the surface temperature $T_s$ using radiative equilibrium based on the incoming solar radiation and obliquity ($\delta=3^\circ$) assuming an albedo of $0.81$. Typical heat losses averaged over the globe are $\mathcal{H}_{\mathrm{cond}}$= $50$~mW/m$^2$, broadly consistent with observations \citep{Tajeddine-Soderlund-Thomas-et-al-2017:true}.

  \underline{Ice-ocean fluxes}

    The interaction between ocean and ice is simulated using MITgcm's ``shelf-ice'' package \citep{Losch-2008:modeling, Holland-Jenkins-1999:modeling}. We turn on the ``boundary layer'' option to avoid possible numerical instabilities induced by an ocean layer which is too thin. The code is modified to account for a gravitational acceleration that is very different from that on earth, the temperature dependence of heat conductivity, and the meridional variation of tidal heating generated inside the ice shell and ice surface temperature. In the description that follows, we begin by introducing the shelf-ice parameterization in a fully coupled ocean-ice system and then make simplifications that fit our goal here. 
    
    Following \citet{Kang-Bire-Campin-et-al-2020:differing}, the freezing/melting rate of the ice shell is determined by a heat budget for a thin layer of ice at the base\footnote{This choice is supported by the fact that most tidal heating is generated close to the ocean-ice interface \citep{Beuthe-2018:enceladuss}.}. The budget involves three terms: the heat transmitted upward by the ocean $\mathcal{H}_{\mathrm{ocn}}$, the heat loss through the ice shell due to heat conduction $\mathcal{H}_{\mathrm{cond}}$ (Eq.\ref{eq:H-cond}), and the tidal heating generated inside the ice shell $\mathcal{H}_{\mathrm{ice}}$ (Eq.\ref{eq:H-tide}). Following the idea of \citet{Holland-Jenkins-1999:modeling} and \citet{Losch-2008:modeling}, the continuity of heat flux and salt flux through the ``boundary layer'' gives,
  \begin{eqnarray}
    &~&\mathcal{H}_{\mathrm{ocn}}-\mathcal{H}_{\mathrm{cond}}+\mathcal{H}_{\mathrm{ice}}=-L_fq-C_p(T_{\mathrm{ocn-top}}-T_b)q\label{eq:boundary-heat}\\
&~&\mathcal{F}_{\mathrm{ocn}}=-S_bq-(S_{\mathrm{ocn-top}}-S_b)q, \label{eq:boundary-salinity}   
  \end{eqnarray}
  where $T_{\mathrm{ocn-top}}$ and $S_{\mathrm{ocn-top}}$ denote the temperature and salinity in the top grid of the ocean\footnote{When model resolution is smaller than the boundary layer thickness, the salinity below the upper-most grid cell also contributes to $T_{\mathrm{ocn-top}}$ and $S_{\mathrm{ocn-top}}$.}, $S_b$ denotes the salinity in the ``boundary layer'', and $q$ denotes the freezing rate in $kg/m^2/s$. $C_p=4000$~J/kg/K is the heat capacity of the ocean, $L_f=334000$~J/kg is the latent heat of fusion of ice.
  
$\mathcal{H}_{\mathrm{ocn}}$ and $\mathcal{F}_{\mathrm{ocn}}$ in Eq.\ref{eq:boundary-heat} can be written as
  \begin{eqnarray}
    \mathcal{H}_{\mathrm{ocn}}&=&C_p(\rho_{0}\gamma_T-q)(T_{\mathrm{ocn-top}}-T_b),\label{eq:T-ocn}\\
    \mathcal{F}_{\mathrm{ocn}}&=&(\rho_{0}\gamma_S-q)(S_{\mathrm{ocn-top}}-S_b) \label{eq:S-ocn}
  \end{eqnarray}
  where $\gamma_T=\gamma_S=10^{-5}$~m/s are the exchange coefficients for temperature and salinity, and $T_b$ denotes the temperature in the ``boundary layer''. The terms associated with $q$ are the heat/salinity change induced by exchange of water between the upper grid box of the ocean model and the ``boundary layer'', where melting and freezing occur. $T_b=T_f(S_b,P)$ is given by the freezing point at pressure $P$ and salinity $S_b$ (see Eq.\ref{eq:freezing-point}).

  In a fully-coupled system, we would solve for $S_b$ and $q$ from Eq.~(\ref{eq:boundary-heat})-(\ref{eq:S-ocn}). However, if we allow freezing and melting of ice and the ocean circulation to feedback onto one-another, the positive feedback between them renders it difficult to find consistent solutions. We therefore cut off this feedback loop by setting the freezing rate $q$ to that which is required to sustain the prescribed ice sheet geometry (details can be found in the next section, ice flow model), whilst allowing a heating term to balance the heat budget (Eq.\ref{eq:boundary-heat}). This also simplifies the calculation of the T/S tendencies of the upper-most ocean grid. The salinity tendency follows directly from  Eq.~(\ref{eq:boundary-salinity}) as
  \begin{equation}
    \frac{dS_{\mathrm{ocn-top}}}{dt}=\frac{-\mathcal{F}_{\mathrm{ocn}}}{\rho_{w0}\delta z}=\frac{qS_{\mathrm{ocn-top}}}{\rho_{w0}\delta z}
  \end{equation}
  
 and the temperature tendency can be approximated as:
  \begin{eqnarray}
    \frac{dT_{\mathrm{ocn-top}}}{dt}&=&\frac{1}{\delta z}(\gamma_T-q)(T_{\mathrm{f,ocn-top}}-T_{\mathrm{ocn-top}}),\label{eq:T-tendency}
  \end{eqnarray}
  where we 
  replaced the boundary layer freezing temperature $T_b=T_f(S_b,P)$ in Eq.~(\ref{eq:T-ocn}) with $T_{\mathrm{f,ocn-top}}=T_f(S_{\mathrm{ocn-top}}, P)$, i.e. the freezing temperature determined by the upmost ocean grid salinity and pressure. According to Eq. (\ref{eq:S-ocn}), and using that $|q|\lesssim 10^{-7}$~kg/$m^2$/s is orders of magnitude smaller than $\rho_0\gamma_S=0.01$~kg/m$^2$/s, the difference between $S_b$ and $S_{\mathrm{ocn-top}}$ can be estimated as $\mathcal{F}_{\mathrm{ocn}}/(\rho_0\gamma_S)=qS_{\mathrm{ocn-top}}/(\rho_0\gamma_S)$. Even in the saltiest scenario we consider here, $|S_b-S_{\mathrm{ocn-top}}|$ does not exceed $0.0004$~psu, and the associated freezing point change is lower than $10^{-5}$K.
  Readers interested in the formulation of a freely evolving ice-water system are referred to the method section of \citet{Kang-Bire-Campin-et-al-2020:differing} and \citet{Losch-2008:modeling}.
  
  In addition to the above conditions on temperature and salinity, the tangential velocity is relaxed back to zero at a rate of $\gamma_M=10^{-3}$m/s at the upper and lower boundaries.

 \subsection{Ice flow model}

  We prescribe $q$ using the divergence of the ice flow, assuming the ice sheet geometry is in equilibrium. We use an upside-down land ice sheet model following \citet{Ashkenazy-Sayag-Tziperman-2018:dynamics}. The ice flows down its thickness gradient, driven by the pressure gradient induced by the spatial variation of the ice top surface, somewhat like a second order diffusive process. At the top, the speed of the ice flow is negligible because the upper part of the shell is so cold and hence rigid; at the bottom, the vertical shear of the ice flow speed vanishes, as required by the assumption of zero tangential stress there. This is the opposite to that assumed in the land ice sheet model. In rough outline, we calculate the ice flow using the expression below obtained through repeated vertical integration of the force balance equation (the primary balance is between the vertical flow shear and the pressure gradient force), using the aforementioned boundary conditions to arrive at the following formula for ice transport $\mathcal{Q}$,
\begin{equation}
  \mathcal{Q}(\phi)= \mathcal{Q}_0H^3(\partial_\phi H/a) \label{eq:ice-flow}
\end{equation}
where
\begin{equation}
\mathcal{Q}_0=\frac{2(\rho_0-\rho_i)g}{\eta_{m}(\rho_0/\rho_i)\log^3\left(T_f/T_s\right)}\int_{T_s}^{T_f}\int_{T_s}^{T(z)}\exp\left[-\frac{E_{a}}{R_{g} T_{f}}\left(\frac{T_{f}}{T'}-1\right)\right]\log(T')~\frac{dT'}{T'}~\frac{dT}{T}.\nonumber 
\end{equation}
Here, $\phi$ denotes latitude, $a$ and $g$ are the radius and surface gravity of the moon, $T_s$ and $T_f$ are the temperature at the ice surface and the water-ice interface (equal to the local freezing point, Eq.~\ref{eq:freezing-point}), and $\rho_i=917$~kg/m$^3$ and $\rho_0$ are the ice density and the reference water density. $E_a=59.4$~kJ/mol is the activation energy for diffusion creep, $R_g=8.31$~J/K/mol is the gas constant and $\eta_{m}$ is the ice viscosity at the freezing point. The latter has considerable uncertainty \citep[$10^{13}$-$10^{16}$~Pa$\cdot$s][]{Tobie-Choblet-Sotin-2003:tidally} and we here set $\eta_{m}=10^{14}$~Pa$\cdot$s, unless otherwise mentioned.

In steady state, the freezing rate $q$ must equal the divergence of the ice transport, thus:
\begin{equation}
    q=-\frac{1}{a\cos\phi}\frac{\partial}{\partial \phi} (Q\cos\phi).
    \label{eq:freezing-rate}
\end{equation}
As shown by the dashed curve in Fig.1b of the main text, ice melts at high latitudes and forms at low latitudes at a rate of a few kilometers every million years. A more detailed description of the ice flow model can be found in \citet{Kang-Flierl-2020:spontaneous} and \citet{Ashkenazy-Sayag-Tziperman-2018:dynamics}. 

\subsection{Model of tidal dissipation in the ice shell}
\label{sec:tidal-dissipation-model}

The ice shell of icy moons is periodically deformed by tidal forcing and the resulting strains in the ice sheet produce heat. We follow \citet{Beuthe-2019:enceladuss} to calculate the implied dissipation rate. Instead of repeating the whole derivation here, we only briefly summarize the procedure and present the final result. Unless otherwise stated, parameters are the same as assumed in \citet{Kang-Flierl-2020:spontaneous}.

Tidal dissipation consists of three components \citep{Beuthe-2019:enceladuss}: a membrane mode $\mathcal{H}_{\mathrm{ice}}^{\mathrm{mem}}$ due to the extension/compression and tangential shearing of the ice membrane, a mixed mode $\mathcal{H}_{\mathrm{ice}}^{mix}$ due to vertical shifting, and a bending mode $\mathcal{H}_{\mathrm{ice}}^{bend}$ induced by the vertical variation of compression/stretching. Following \citet{Beuthe-2019:enceladuss}, we first assume the ice sheet to be completely flat. By solving the force balance equation, we obtain the auxiliary stress function $F$, which represents the horizontal displacements, and the vertical displacement $w$. The dissipation rate $\mathcal{H}_{\mathrm{ice}}^{\mathrm{flat,x}}$ (where $x=\{\mathrm{mem},\mathrm{mix},\mathrm{bend}\}$ ) can then be written as a quadratic form of $F$ and $w$. In the calculation, the ice properties are derived assuming a globally-uniform surface temperature of 60K and a melting viscosity of $5\times10^{13}$~Pa$\cdot$s. 

Ice thickness variations are accounted for by multiplying the membrane mode dissipation $\mathcal{H}_{\mathrm{ice}}^{\mathrm{flat,mem}}$, by a factor that depends on ice thickness. The membrane mode is the only mode which is amplified in thin ice regions (see \citet{Beuthe-2019:enceladuss}). This results in the expression:
\begin{equation}
  \label{eq:H-tide}
  \mathcal{H}_{\mathrm{ice}}=(H/H_0)^{p_\alpha}\mathcal{H}_{\mathrm{ice}}^{\mathrm{flat,mem}}+\mathcal{H}_{\mathrm{ice}}^{\mathrm{flat,mix}}+\mathcal{H}_{\mathrm{ice}}^{\mathrm{flat,bend}},
\end{equation}
where $H$ is the prescribed thickness of the ice shell as a function of latitude and $H_0$ is the global mean of $H$. Since thin ice regions deform more easily and produce more heat, $p_\alpha$ is negative. Because more heat is produced in the ice shell, the overall ice temperature rises, which, in turn, further increases the mobility of the ice and leads to more heat production (the rheology feedback).


The tidal heating profile corresponding to $p_\alpha=-1.5$ is the red solid curve plotted in Fig.1c of the main text.

\section{Idealized Ice evolution model.}
Here we provide a brief overview for the idealized model used to evolve the ice shell of Enceladus and Europa. Interested readers are referred to \citep{Kang-Flierl-2020:spontaneous} and its supplementary material for more detail.

In this model, ice shell thickness $H$ changes over time due to the melting induced by the tidal heating $\mathcal{H}_{\mathrm{ice}}$ (given by Eq.~\ref{eq:H-tide}), the down-gradient ice flow $\mathcal{Q}$ (given by Eq.~\ref{eq:ice-flow}), the heat loss to space by conduction $\mathcal{H}_{\mathrm{cond}}$ (given by Eq.~\ref{eq:H-cond}), the crack-induced cooling $\mathcal{H}_{\mathrm{crack}}$ in places where ice is sufficiently thin, and heat transmitted upward by the ocean $\mathcal{H}_{\mathrm{ocn}}$. The ice thickness tendency can be symbolically expressed as follows,
\begin{equation}
  \label{eq:ice-evolution-A}
  \frac{dH}{dt}=\frac{\mathcal{H}_{\mathrm{cond}}(H)-\mathcal{H}_{\mathrm{ice}}(H)-\mathcal{H}_{\mathrm{ocn}}}{L_f\rho_i } + \frac{1}{a\sin\phi}\partial_\phi \left(\sin\phi \mathcal{Q}(H)\right),
\end{equation}
where $L_f$ and $\rho_i$ are the latent heat of freezing and density of ice, $a$ is the moon's radius and $\phi$ denotes latitude.
Physical constants and parameters for Enceladus and Europa are stated in Table.\ref{tab:parameters-ice}. $\mathcal{H}_{\mathrm{ice}}$ is polar-amplified, and as a result, the polar ice shell tends to be thinner, which in turn increases the heat production over the pole (see Eq.~\ref{eq:H-tide}). The tendency for ice thickness variations to increase due to the rheology feedback will be balanced by the rapid heat loss through thin ice (Eq.~\ref{eq:H-cond}), the transport by ice flow (Eq.~\ref{eq:ice-flow}) and the ocean heat transport. An additional heat sink is activated only when the ice thickness is less than $H_{\mathrm{crack}}=3$~km to prevent further melting, and crudely represents the effect of cracks and geysers that carry the extra heat away. At all times, the global tidal dissipation $\mathcal{H}_{\mathrm{ice}}$ is scaled to exactly balance the instantaneous conductive heat loss $\mathcal{H}_{\mathrm{cond}}$. By so doing, the rheology feedback and thus the ice thickness variation are maximized. Throughout the integration, the global mean ice thickness is fixed at $H_0=20$~km.

The ocean-ice heat exchange is a new process we introduced. Inspired by the conceptual model, the heat flux coming from the ocean is parameterized by Eq.~19 in the main text. Definitions and values for other parameters can be found in Table.~\ref{tab:parameters} and Table~\ref{tab:parameters-ice}.

The initial condition is set as follows
\begin{equation}
  \label{eq:ice-evolution-IC}
  H(0)=H_0-H_2P_2(\sin\phi)-H_1P_1(\sin\phi),
\end{equation}
where $H_0=20$~km, $H_2=3$~km and $H_1=1$~km. $P_1$ and $P_2$ are the first and second order Legendre Polynomials. 


\begin{table*}[hptb!]
  
  \centering
  \begin{tabular}{lll}
    Symbol & Name & Definition/Value\\
    \hline
    \multicolumn{3}{c}{Physical constants}\\
    \hline
    $L_f$ & fusion energy of ice & 334000~J/kg\\
    $C_p$ & heat capacity of water & 4000~J/kg/K\\
    $T_f(S,P)$ & freezing point & Eq.\ref{eq:freezing-point}\\
    $\rho_i$ & density of ice & 917~kg/m$^3$ \\
    $\rho_w$ & density of the ocean & 'MDJWF' eq. of state \citep{McDougall-Jackett-Wright-et-al-2003:accurate} \\
    $\alpha$ & thermal expansion coeff. &  $-\partial (\rho/\rho_0)/\partial T$ \\
    $\beta$ & saline contraction coeff. &  $\partial (\rho/\rho_0)/\partial S$\\
    $\kappa_0$ & conductivity coeff. of ice & 651~W/m\\
    $p_\alpha$ & ice dissipation amplification factor & -2 $\sim$ -1 \\
    $\eta_{m}$ & ice viscosity at freezing point & 10$^{14}$~Ps$\cdot$s\\
    \hline
    \multicolumn{3}{c}{Parameters for the size-dependence experiments}\\
    \hline
    $a$ & radius & 150,\ 250,\ 500,\ 1000,\ 1500,\ 2500~km\\
    $g_0$ & surface gravity & Eq.XY\\
    $\delta$ & obliquity & 3.1$^\circ$\\
    $H_0$ & global mean ice thickness & 20~km  \\
    $H_2$ & equator-to-pole ice thickness variation & 3~km\\
    $D$ & global mean ocean depth& 56~km \\
    $\Omega$ & rotation rate & 2.05$\times$10$^{-5}$~s$^{-1}$\\
    $\bar{T_s}$ & mean surface temperature& 110K\\
    $S_0$ & mean ocean salinity & 10,\ 60~psu\\
    $c$ & proportion of heat generated in the core & 0\\
    \hline
    \multicolumn{3}{c}{Parameters used in the thickness varying simulations}\\
    \hline
    $a$ & radius & 1561~km\\
    $g_0$ & surface gravity & 1.315~m/s$^2$\\
    $\delta$ & obliquity & 3.1$^\circ$\\
    $H_0$ & global mean ice thickness  &  15~km \citep{Hand-Chyba-2007:empirical} \\
    $H_2$ & equator-to-pole ice thickness variation & 0.1,\ 0.2,\ 0.5,\ 1,\ 2~km\\
    $D$ & global mean ocean depth & 85~km \citep{Hand-Chyba-2007:empirical} \\
    $\Omega$ & rotation rate & 2.05$\times$10$^{-5}$~s$^{-1}$\\
    $\bar{T_s}$ & mean surface temperature & 110K \\
    $S_0$ & mean ocean salinity & 60~psu \citep{Hand-Chyba-2007:empirical}\\
    $c$ & proportion of heat generated in the core & 0\%\\
    \hline
    \multicolumn{3}{c}{Other parameters in the ocean model}\\
    \hline
    $\nu_h$ & horizontal viscosity & $\frac{a}{150~\mathrm{km}}$~m$^2$/s\\
    $\nu_v$ & vertical viscosity & 1~m$^2$/s\\
    $\tilde{\nu}_h,\ \tilde{\nu}_v$ & bi-harmonic hyperviscosity & $\frac{a}{150~\mathrm{km}}\times$10$^8$~m$^4$/s\\ 
    $\kappa_h,\ \kappa_v$ & horizontal/vertical diffusivity & 0.001~m$^2$/s\\
    $(\gamma_T,\ \gamma_S,\ \gamma_M)$ & water-ice exchange coeff. for T, S \& momentum & (10$^{-5}$, 10$^{-5}$, 10$^{-4}$)~m/s\\
    $g$ & gravity in the ocean interior & Eq.\ref{eq:g-z}\\
    $P_0$ & reference pressure & $\rho_ig_0H_0$ \\
    $T_0$ & reference temperature & $T_f(S_0,P_0)$ \\
    $H$ & ice shell thickness & Eq.\ref{eq:Hice}\\
    $\mathcal{H}_{\mathrm{cond}}$ & conductive heat loss through ice & Eq.\ref{eq:H-cond}\\
    $\mathcal{H}_{\mathrm{ice}}$ & tidal heating produced in the ice & Eq.\ref{eq:H-tide} \\
    \hline
     \end{tabular}
  \caption{Model parameters used in the ocean general circulation model and the conceptual model. }
  \label{tab:parameters}
  
\end{table*}

\begin{table*}[hptb!]
  
  \centering
  \begin{tabular}{lll}
    Symbol & Name & Definition/Value\\
    \hline
    $p_\alpha$ & ice dissipation amplification factor & -1.5\\
\hline
    \multicolumn{3}{c}{Parameters for Enceladus}\\
    \hline
    $a$ & radius & 252~km\\
    $g_0$ & surface gravity & 0.113~m/s$^2$\\
    $\delta$ & obliquity & 27$^\circ$\\
    $H_0$ & global mean ice thickness &  20~km \citep{Hemingway-Mittal-2019:enceladuss} \\
    $H_2$ & initial equator-to-pole ice thickness variation & 3~km\\
    $H_1$ & initial hemispherical asymmetry & 1~km\\
    $\bar{T_s}$ & mean surface temperature & 59K \\
    \hline
    \multicolumn{3}{c}{Parameters for Europa}\\
    \hline
    $a$ & radius & 1561~km\\
    $g_0$ & surface gravity & 1.315~m/s$^2$\\
    $\delta$ & obliquity & $3.1^\circ$\\
    $H_0$ & global mean ice thickness &  20~km \\
    $H_2$ & initial equator-to-pole ice thickness variation & 3~km\\
    $H_1$ & initial hemispherical asymmetry & -1~km\\
    $\bar{T_s}$ & mean surface temperature & 110K \\
    \hline
     \end{tabular}
  \caption{Parameters used in the ice evolution model. }
  \label{tab:parameters-ice}
  
\end{table*}

\begin{figure*}[htp!]
    \centering
    \includegraphics[page=5,width=0.9\textwidth]{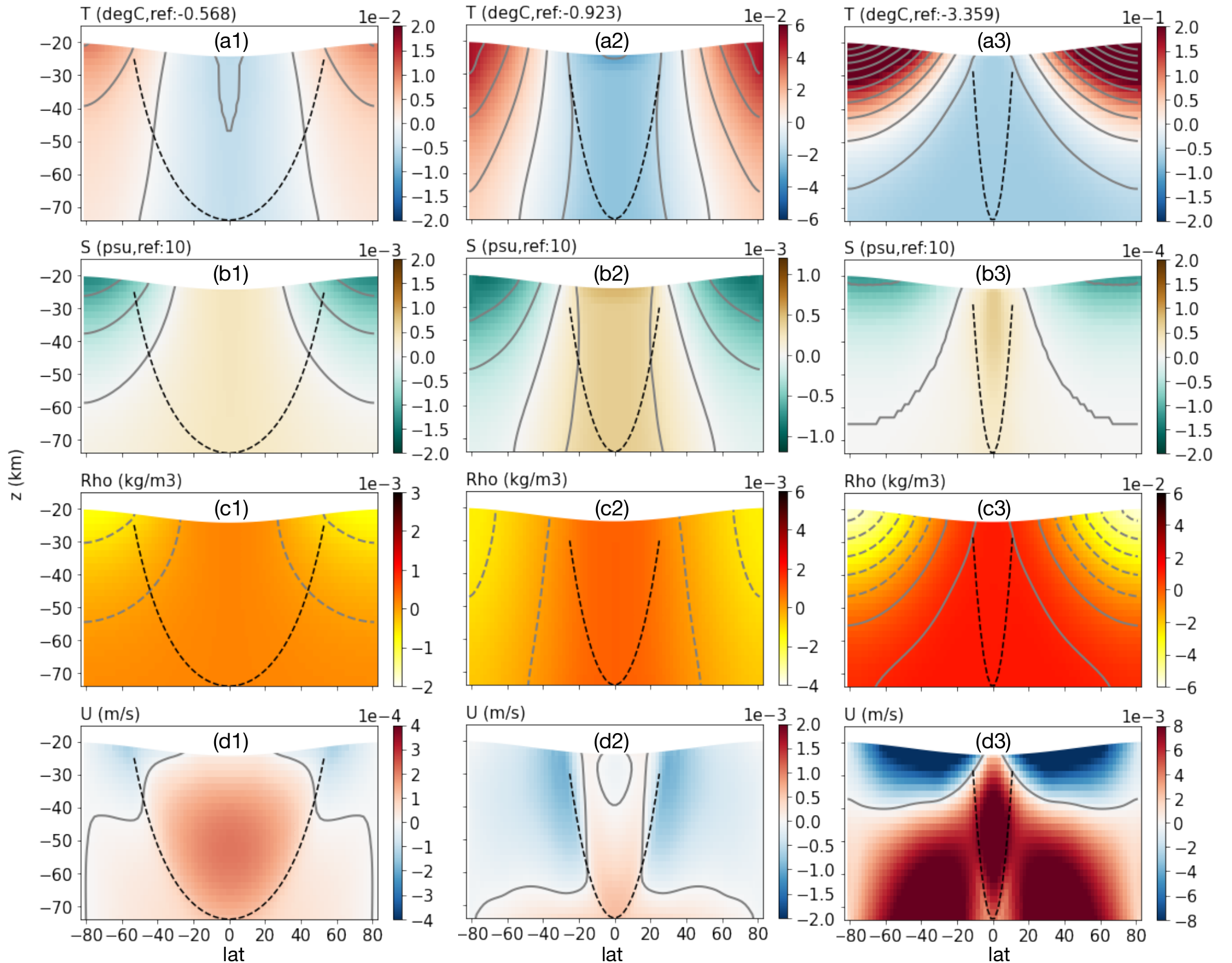}
    
    \caption{\small{Ocean circulation and thermodynamic state driven by the glacial melt and under-ice temperature distributions for icy moons with different sizes. Rows form top to bottom show temperature $T$, salinity $S$, density $\Delta\rho$ and zonal flow speed $U$. Columns from left to right show results for a small icy moon with 150~km radius, a medium icy moon with 500~km radius and a large icy moon with 2500~km radius, respectively. Ocean salinity is assumed to be 10~psu and the friction rate $\gamma=10^{-3}$~m/s. Black dashed lines mark the position of the tangent cylinder that touches the solid core at the equator.}}
    
    \label{fig:T-S-Rho-Psi-S10}
  \end{figure*}

  \begin{figure*}[htp!]
    \centering
    \includegraphics[page=6,width=0.9\textwidth]{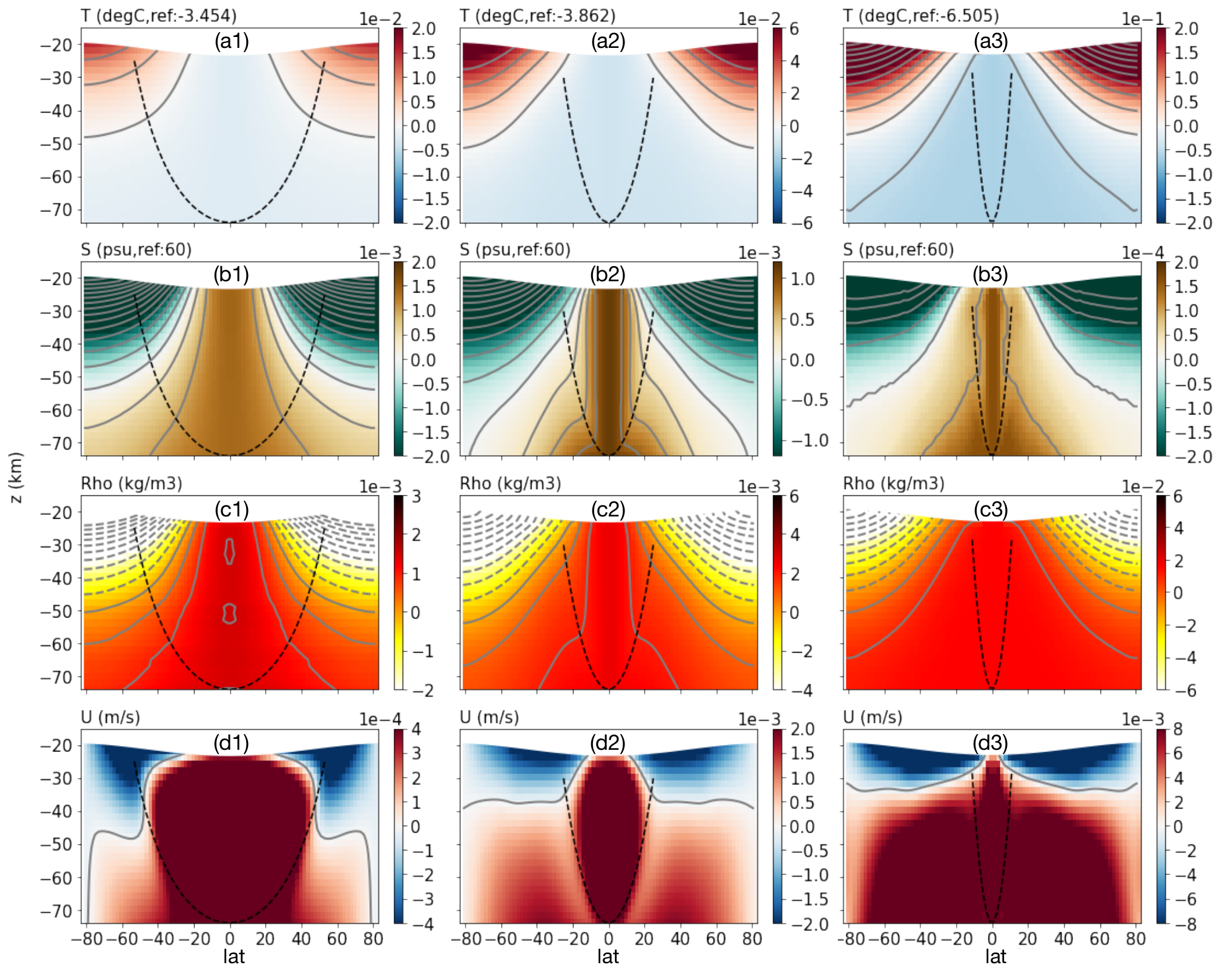}
    \caption{\small{Same as Fig.~\ref{fig:T-S-Rho-Psi-S10} except that ocean salinity is set to 60~psu.}}
    \label{fig:T-S-Rho-Psi-S60}
  \end{figure*}

  \begin{figure*}
    \centering \includegraphics[page=13,width=0.9\textwidth]{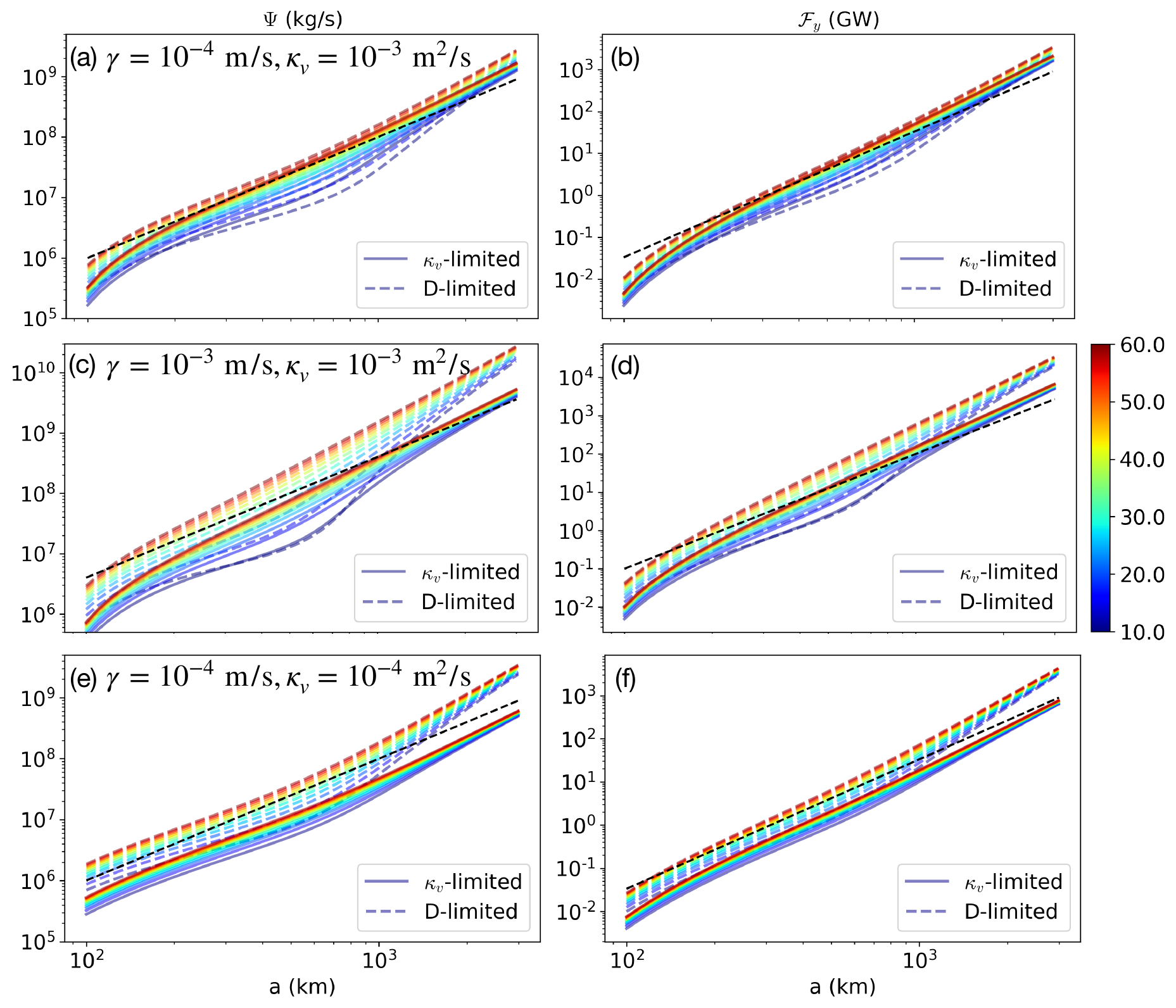}
    \caption{\small{The overturning streamfunction $\Psi$ (left) and the meridional heat transport $\mathcal{F}_y$ (right) predicted by the $\kappa_v$-limited scaling (solid lines) and the $D$-limited scaling (dashed lines). Panels (a,b) are for the default experiment, panels (c,d) are for the sensitivity test with 10 times larger friction rate ($\gamma=10^{-3}$~m/s), and panels (e,f) are for the sensitivity test with 10 times smaller diffusivity ($\kappa_v=10^{-4}$~m$^2$/s). Whichever scaling leads to a lower prediction for $\Psi$ is taken as the final prediction. As can be seen, fresh oceans with intermediate satellite size tend to be $D$-limited, while other cases tend to be $\kappa_v$-limited.}}
    \label{fig:Dlimit-kappalimit}
  \end{figure*}

  \begin{figure*}
    \centering \includegraphics[page=7,width=0.9\textwidth]{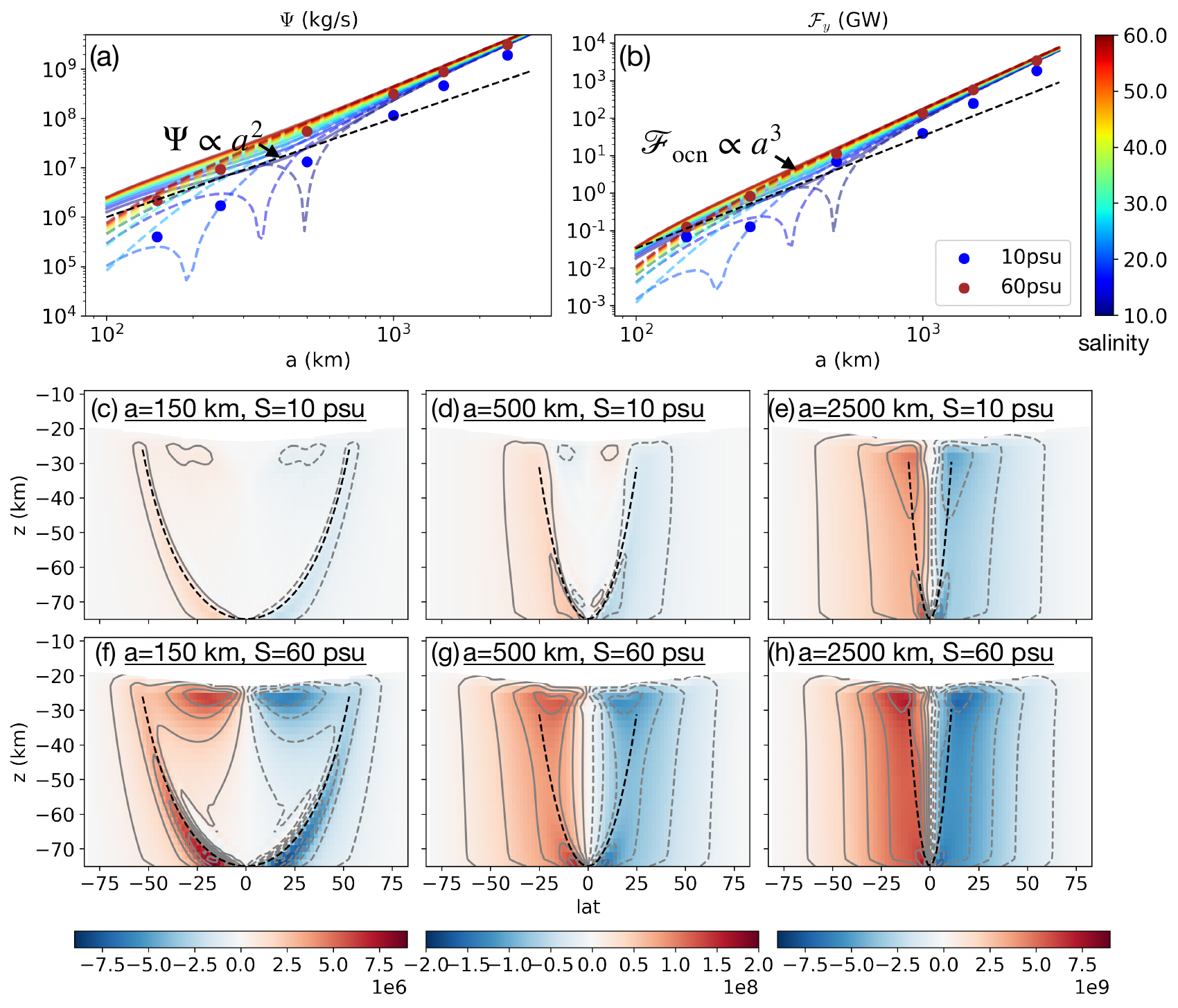}
    \caption{\small{As Fig.~2 in the main text except for $\gamma=10^{-3}$~m/s.}}
    \label{fig:scaling-circulation-gamma1e-3}
  \end{figure*}

      \begin{figure*}
    \centering
    \includegraphics[page=8,width=0.9\textwidth]{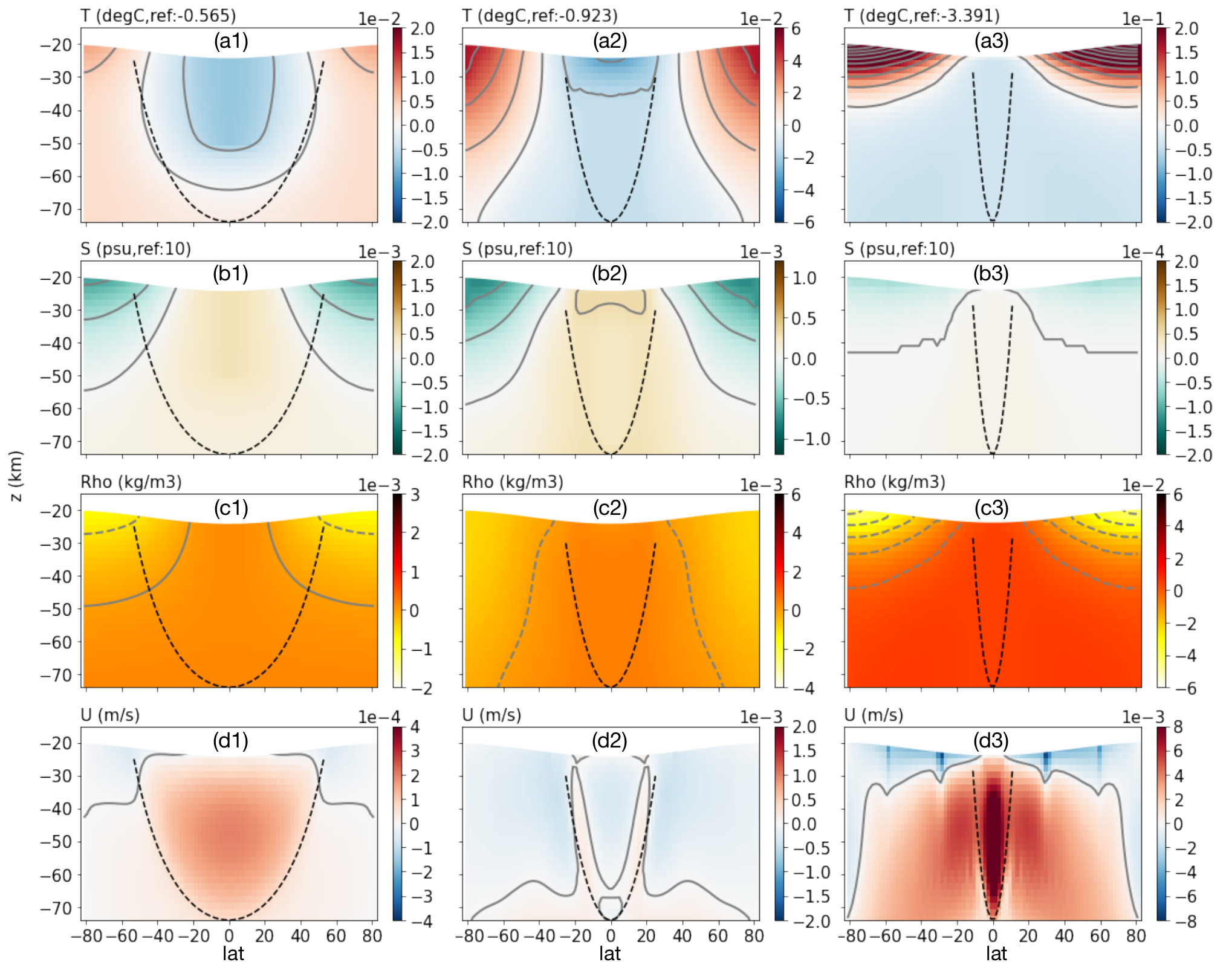}
    \caption{\small{As Fig.~\ref{fig:T-S-Rho-Psi-S10} except that $\gamma$ is set to $10^{-3}$~m/s.}}
    \label{fig:T-S-Rho-Psi-S10-gamma1e-3}
  \end{figure*}

  \begin{figure*}
    \centering
    \includegraphics[page=9,width=0.9\textwidth]{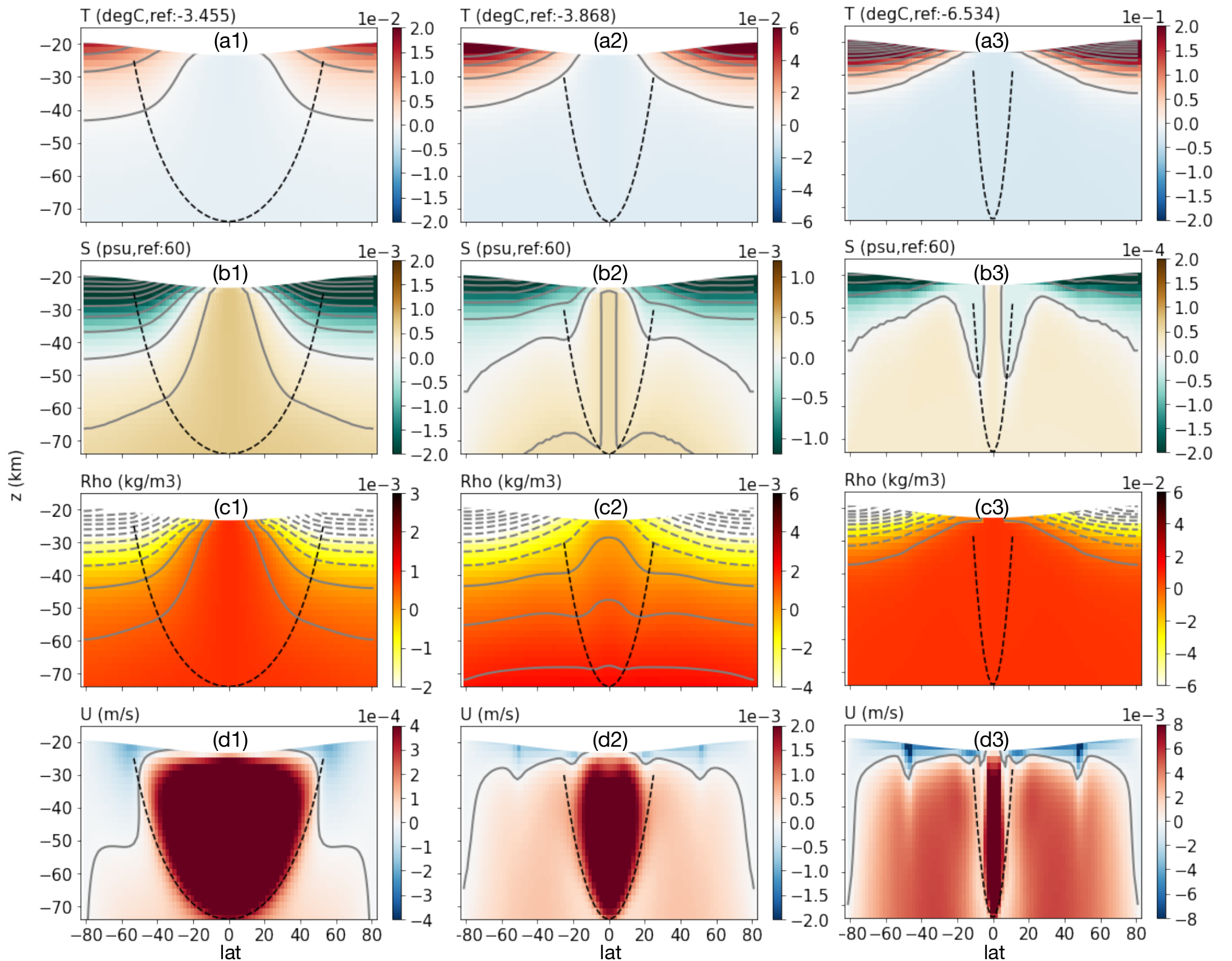}
    \caption{\small{As Fig.~\ref{fig:T-S-Rho-Psi-S60} except that $\gamma$ is set to $10^{-3}$~m/s.}}
    \label{fig:T-S-Rho-Psi-S60-gamma1e-3}
  \end{figure*}

    \begin{figure*}
    \centering \includegraphics[page=10,width=0.9\textwidth]{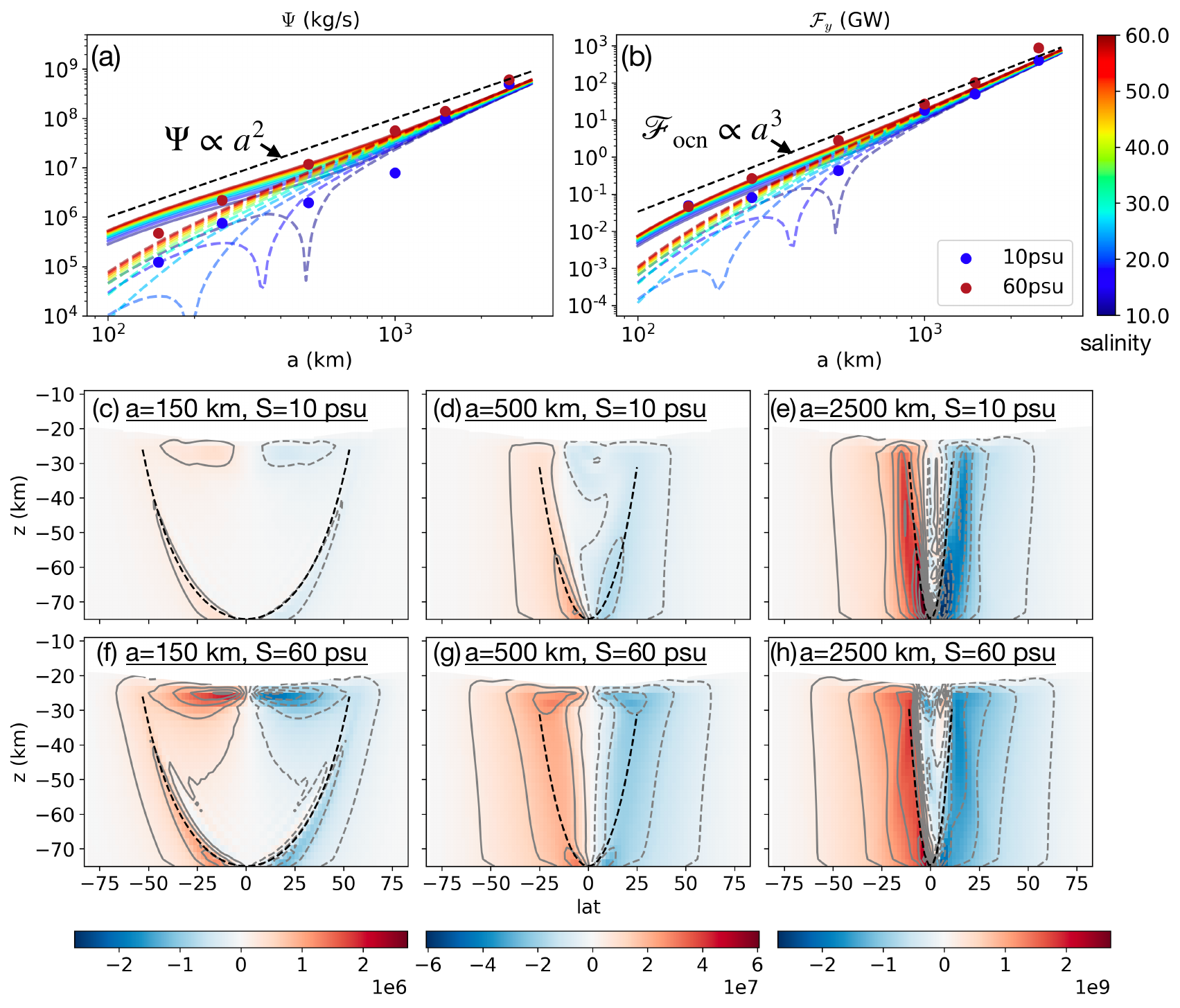}
    \caption{\small{As Fig.~2 in the main text except for $\kappa_v=10^{-4}$~m$^2$/s.}}
    \label{fig:scaling-circulation-diff1e-4}
  \end{figure*}

      \begin{figure*}
    \centering
    \includegraphics[page=11,width=0.9\textwidth]{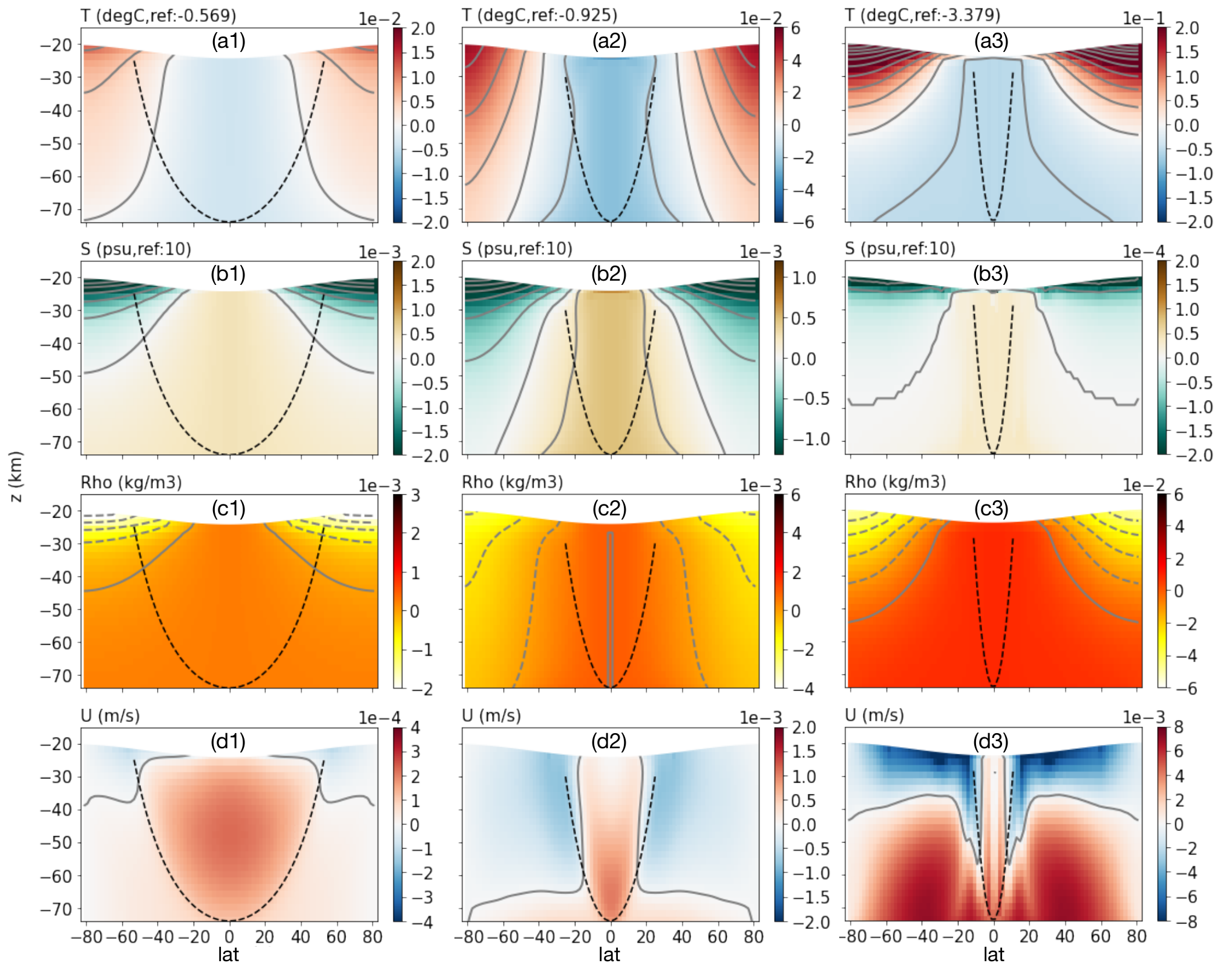}
    \caption{\small{As Fig.~\ref{fig:T-S-Rho-Psi-S10} except that $\kappa_v=10^{-4}$~m$^2$/s.}}
    \label{fig:T-S-Rho-Psi-S10-diff1e-4}
  \end{figure*}

  \begin{figure*}
    \centering
    \includegraphics[page=12,width=0.9\textwidth]{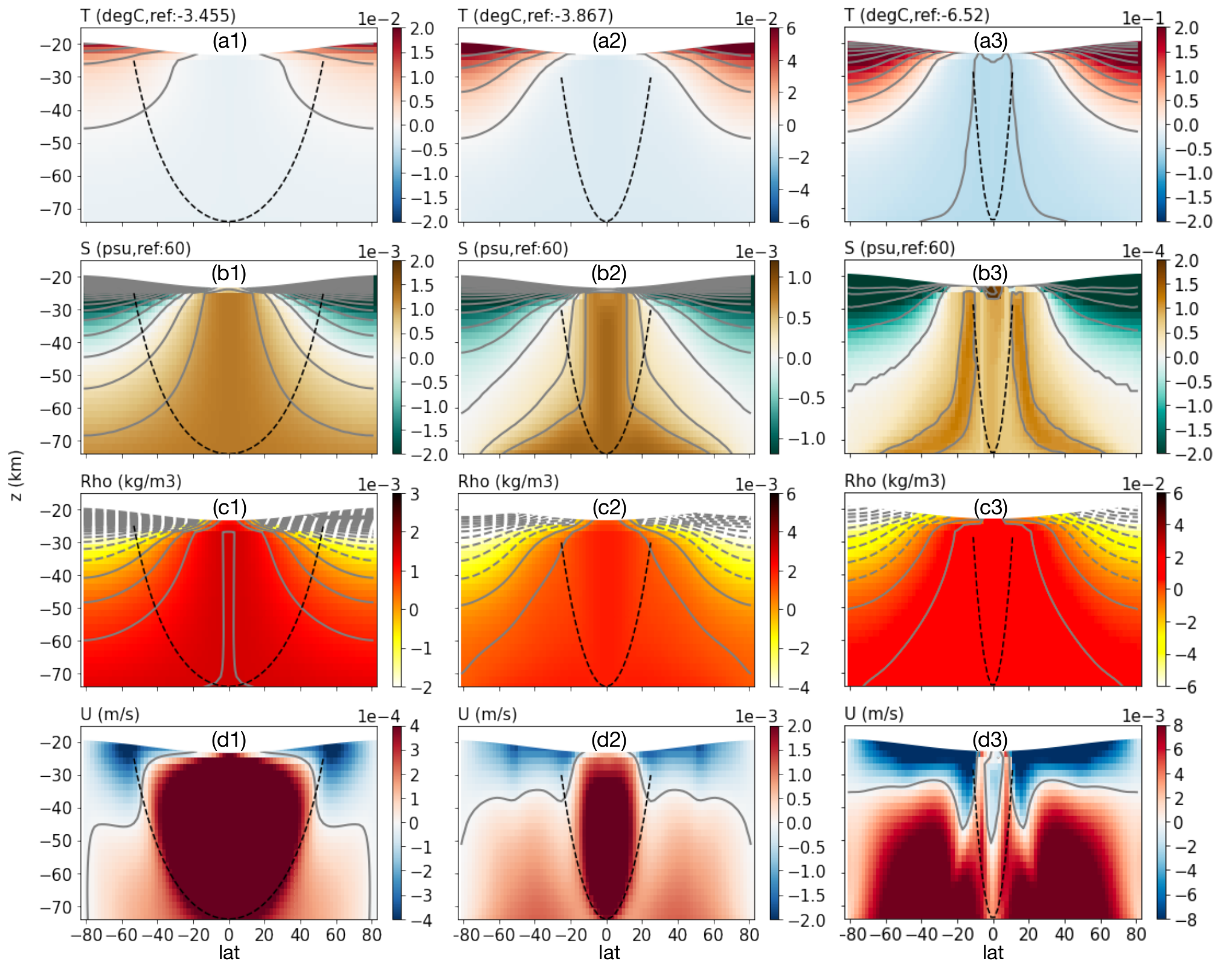}
    \caption{\small{As Fig.~\ref{fig:T-S-Rho-Psi-S60} except that $\kappa_v=10^{-4}$~m$^2$/s.}}
    \label{fig:T-S-Rho-Psi-S60-diff1e-4}
  \end{figure*}

  \begin{figure*}
    \centering
    \includegraphics[page=14,width=0.6\textwidth]{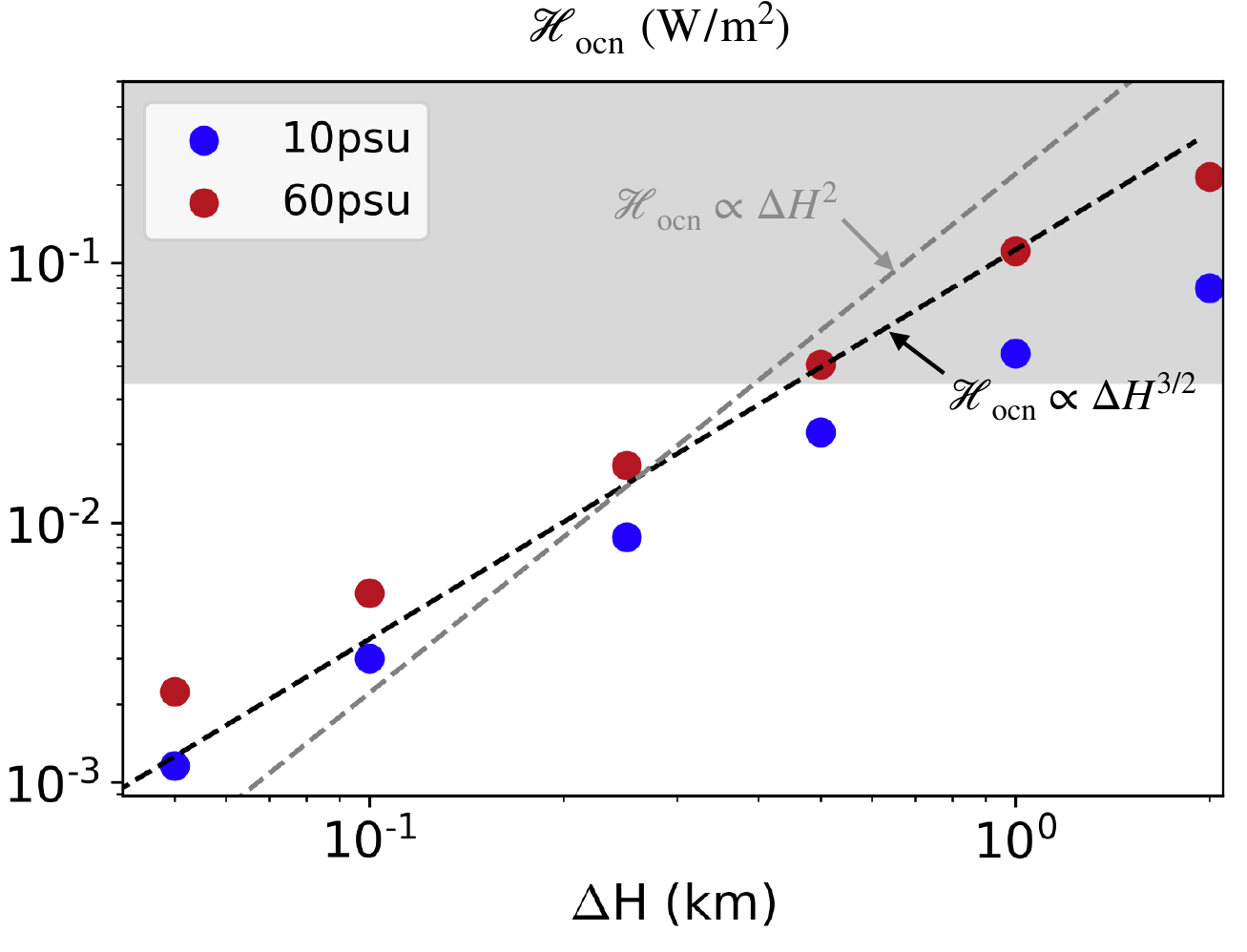}
    \caption{\small{As Fig.~3 in the main text except for $\gamma=10^{-3}$~m/s. }}
    \label{fig:heat-transport-europa-varydH-gamma1e-3}
  \end{figure*}

\bibliography{export}{}
\bibliographystyle{aasjournal}



\end{document}